# The role of integrability in a large class of physical systems

D. H. Delphenich [†]
Kettering, OH 45440 USA

**Abstract.** A large class of physical systems involves the vanishing of a 1-form on a manifold as a constraint on the acceptable states. This means that one is always dealing with the Pfaff problem in those cases. In particular, knowing the degree of integrability of the 1-form is often essential, or, what amounts to the same thing, its canonical (i.e., normal) form. This paper consists of two parts: In the first part, the Pfaff problem is presented and discussed in a largely mathematical way, and in the second part, the mathematical generalities thus introduced are applied to various physical models in which the normal form of a 1-form has already been implicitly introduced, such as non-conservative forces, linear non-holonomic constraints, the theory of vortices and equilibrium thermodynamics. The role of integrability in the conservation of energy is a recurring theme.

## Contents

---

[†] Website: neo-classical-physics.info

# INTRODUCTION

Because the fundamental laws of physics, whether one is considering the laws that govern the static – i.e., equilibrium – state of a continuously-distributed physical system or the laws that govern its time evolution, often take the form of differential equations, one must always examine the conditions under which those differential equations are and are not actually integrable.

In many cases, one finds that the integrability of the system in question is simply assumed as a convenient simplifying approximation. For instance, the only way that one can treat a "laboratory frame" as being inertial – i.e., holonomic – is usually by ignoring the fact that any frame that is fixed in the Earth will be rotating with the Earth and is therefore fundamentally anholonomic. Thus, the usual process of defining a coordinate system by first defining a local frame field and then integrating the local frame field into a local coordinate system is based on the assumption that the local frame field is holonomic; i.e., integrable. One then compares the numerical magnitude of the actual anholonomy of the local frame field to some acceptable degree of error to see whether it is safe to ignore the non-integrability of the frame field. For instance, in a typical classroom demonstration of the laws of elementary collisions in the Newtonian approximation, since the time interval in which the collision takes place is a fraction of a second, the Coriolis contribution to the acceleration from the rotation of the Earth has an order of magnitude of $2\omega v$, where $\omega = 73$ $\mu$rad/sec is the angular velocity of the Earth and $v$ is the velocity of the moving object under consideration. Consequently, when the velocities involved are of order 1 m/s and the time of collision is less than a second the position error that is introduced by ignoring the Coriolis contribution is of order 73 microns. However, when a high-velocity rifle bullet is moving closer to 1000 m/s for a second the lateral position error is of order 73 mm or 7.3 cm, which is an appreciable error in that context. Of course, if the length of time of the experiment has the order of magnitude one day, one can clearly observe the non-integrable nature of the Earth-fixed reference frame in the precession of the Foucault pendulum.

One further finds that, in effect, the nature of most physical approximations amounts to assuming that the system in question can be decomposed into an "internal" subsystem and an "external" one. This basic assumption of decomposability is based on the approximation in which the only way that the state of one subsystem can influence the state of the other one is by way of the "signals" that are exchanged through the "interface" between them. At the level of systems of equations, this means that the state variables of the two subsystems are not coupled by the equations. Of course, sometimes this situation is already closely related to the choice of frame in which one is obtaining the state variables, since there might be a "diagonalizing" frame in which the state variables are decoupled, such as one finds with the normal coordinates of a system of linear oscillators. One then must always address the fundamental problem of whether that frame even exists, which often becomes an algebraic problem in which diagonalizability is equivalent to the commutativity of some relevant operators.

Common examples of this sort of decomposability assumption are the test particle approximation, which is closely related to the reduction of the two-body problem to the one-body problem, the external field approximation, and the heat bath approximation in thermodynamics.

The way that this can affect the integrability is often manifested in the nature of the forces that act on a physical system. One finds that the conservative forces tend to be the ones that come from complete mathematical models for the interaction of particles, whereas the non-conservative ones generally represent empirical approximations to more complex situations, such as the interaction of macroscopic materials in the form of friction or viscous drag. Since a force is conservative iff total energy is conserved in the motion that results from the action of that force, one sees that the integrability of the force 1-form is closely related to not only the completeness in the model for the force, but also the conservation of energy in the system. One finds that the conservation of energy can be described by saying that allowable kinematical states must be constrained to lie in integral submanifolds of a Pfaffian equation that relates to the force 1-form.

One then sees how a broad class of physical systems can involve some sort of differential constraint that takes the form of the exterior differential equation $\theta = 0$, where $\theta$ is a 1-form. Thus, the allowable states of the system would have to lie on integral submanifolds of that 1-form, so one needs to investigate the structure of the integral submanifolds.

In mathematics, this class of problems goes back to the work of Johann Friedrich Pfaff [1] in 1814, and is thus referred to as the *Pfaff problem*, while a 1-form is often referred to as a *Pfaffian form*. Since the time of Pfaff, the problem that he had posed attracted the attention of most of the major names of Nineteenth Century mathematics. In particular, notable advances in the understanding of the problem were due to Jacobi [2], Natani [3], Clebsch [4], Grassmann [5], Frobenius [6], Mayer [7], Darboux [8], Lie [9], and Cartan [10]. (One finds excellent summaries of the progress in that problem in the books of Forsythe [11] and Goursat [12].)

In the process, the Pfaff problem was intersecting strongly with other major developments of that century, such as Grassmann's theory of exterior algebra, the Plücker-Klein theory of line geometry, and Lie's treatment of the solutions to systems of differential equations by means of differentiable transformation groups, which drew heavily upon the work of Poisson and Jacobi regarding the Pfaff problem. In the work of Lie, one also finds that the Pfaff problem could benefit greatly from the line geometry that he had learned from Klein.

Since the vanishing of a 1-form at a point of an $n$-dimensional differentiable manifold defines a hyperplane in its tangent space, and the tangent space to an integral submanifold at that point must be contained in that hyperplane, one sees that integral submanifolds can potentially have dimensions that range from one to $n - 1$. The maximum dimension for an integral submanifold that can exist is then called the *degree of integrability* of the Pfaff equation. One finds that any Pfaffian form $\theta$ admits one of two canonical forms, namely, $\mu_i \, dv^j$ or $d\lambda + \mu_i \, dv^j$, where the number of terms in the sum is equal to the codimension of the maximal integral submanifolds. The degree of integrability is then the dimension of the manifold minus that minimum codimension. It was later established that one could find that degree by examining a sequence of differential forms that were defined by exterior products of powers of $d\theta$ and the exterior product of $\theta$ with such powers.

One sees that the expression $d\lambda + \mu_i \, dv^i$, in particular, exhibits $\theta$ as an exact form plus successive inexact correction terms, if one regards exactness as only an approximation to something more physical. For instance, when $\theta$ is a force 1-form, exactness is equivalent to the force being conservative, which is related to the total energy in the system being conserved. Thus, to the extent that conservation of energy is an approximation, the exactness of the force 1-form is also an approximation. As we will see in what follows, this sort of situation is widespread in physical models.

The first part of this paper is then concerned with the mathematical definition and examination of the Pfaff problem. After stating the basic problem and establishing how one finds the degree of integrability and canonical form for a Pfaffian form, we then show that an important application of it is in contact geometry, which has the manifold $J^1(M; \mathbb{R})$ of all 1-jets of differentiable functions on a differentiable manifold $M$ as its prototype, in the same way that symplectic manifolds have the cotangent bundle $T^*(M)$ for their prototype. Indeed, the two spaces are closely related.

Although we shall briefly mention the more general Pfaffian systems, in which one has more than one linearly independent 1-form that all simultaneously vanish on integral manifolds, the difference in the level of complexity between one and more than one Pfaffian form is like the difference between first-order partial differential equations for a single function and systems of first-order partial differential equations in more than one function. Hence, we shall stop short of discussing the integrability problem for Pfaffian systems, which gets one into the Cartan-Kähler theorem, and which deserves a special treatment of its own.

In the second part of the paper, we then apply this general mathematical theory to various examples of mathematical models for physical processes in which one is dealing with a Pfaff form as a fundamental object, such as when one has a differential constraint of Pfaffian type. As we shall see, these examples are quite far-reaching and rooted in fundamental physics, such as the difference between the conservation of energy in a closed system and the balance of energy in an open system. In addition to discussing the work functional and non-conservative force fields, we go on to the way that one can extend the scope of symplectic mechanics from Hamiltonian vector fields, which originate in exact 1-forms, to non-Hamiltonian ones, which come from more general 1-forms. We also show that the integrability of Pfaffian forms is intrinsic to the problems of non-holonomic linear constraints, equilibrium thermodynamics, the vortical flow of fluids, and $U(1)$ gauge field theories, such as electromagnetism.

Although a certain familiarity with the basic notions of differentiable manifolds and differential forms will be assumed, nonetheless, the local expressions for the abstract expressions will usually be given in order to make contact with the more conventional physics treatments of the various subjects.

# I. MATHEMATICAL THEORY OF THE PFAFF EQUATION

**1. The Pfaff problem.** A *Pfaffian form (or Pfaffian expression)* is, quite simply, a differential 1-form $\omega$ on an $n$-dimensional differentiable manifold $M$. Thus, for some local coordinate system $(U, x^i)$ on $M$, $\omega$ will take the form:

$$\omega = \omega_i \, dx^i \, . \tag{1.1}$$

An *integral submanifold (or simply, an integral)* of a Pfaffian form $\omega$ is a differentiable embedding $x: \Pi \to M$, $u \mapsto x(u)$, where $\Pi$ is a $p$-dimensional parameter manifold, such that:

$$x^* \omega = 0. \tag{1.2}$$

Thus, if $(S, u^a)$ is a local coordinate chart on $\Pi$ then the image of $x$ can be expressed by a system of $n$ equations in $p$ variables:

$$x^i = x^i(u^a) \tag{1.3}$$

and the local form of the condition (1.2) that these points must lie on an integral submanifold is obtained by making the replacement:

$$dx^i = \frac{\partial x^i}{\partial u^a} \, du^a \tag{1.4}$$

in (1.1), which then gives the following system of $p$ linear, first-order, partial differential equations for the $n$ functions $x^i$:

$$\omega_i \, \frac{\partial x^i}{\partial u^a} = 0. \tag{1.5}$$

One can define $p$ linearly-independent tangent vectors on $M$ by way of:

$$\mathbf{v}_a = x_* \frac{\partial}{\partial u^a} = \frac{\partial x^i}{\partial u^a} \frac{\partial}{\partial x^i} \qquad (a = 1, \ldots, p) \tag{1.6}$$

so (1.5) takes the form of the set of $p$ linear algebraic equations for vectors that are annihilated by the 1-form $\omega$:

$$0 = \omega(\mathbf{v}) = \omega_i \, v^i \, . \tag{1.7}$$

The solutions to this equation at each point $x \in M$ will represent a hyperplane Ann[$\omega$]$_x$ in its tangent space $T_x M$, so the solution to the *algebraic* equation $\omega = 0$ is a vector sub-bundle Ann[$\omega$] in $T(M)$ whose fibers have corank one. Thus, one can decompose the process of solving the system of equations (1.5) for $x^i$ into two steps:

1.  Finding the algebraic solution Ann[$\omega$], and, in particular, $p$ linearly independent local vector fields $\mathbf{v}_a(x) = v_a^i(x^j)\partial/\partial x^i$ that all lie in Ann[$\omega$].

2.  Solving the system of first-order, linear partial differential equations for $x^i$:

$$\frac{\partial x^i}{\partial u^a} = v_a^i(x^1(u^b),\dots,x^n(u^b))\,. \tag{1.8}$$

This can also be expressed in terms of $n$ local differential forms on $\Pi$ by multiplying both sides of (1.8) by $du^a$, which gives $n$ equations:

$$x^* dx^i = v^i \qquad (i = 1, \dots, n). \tag{1.9}$$

Since the left-hand side of these equations are exact 1-form, it is necessary that the right-hand sides be exact, as well. It is then necessary that the 1-forms $v^i$ must be closed, as well. If $M$ is not simply-connected then this necessary condition is only locally sufficient.

One notes that there will generally be a range of possible dimensions $p$. In particular, since (1.8) becomes a system of ordinary differential equations when $p = 1$, as long as the functions $v_a^i(u)$ are assumed to be continuously differentiable functions of $x^j$, there will always be local solutions; i.e., local flows. At the other end of the scale, $p$ can certainly not be more than $n - 1$.

The maximum such dimension $p$ is called the *degree of integrability* of $\omega$. The *Pfaff problem* can then be given the concise modern form ([1]):

> *For a given 1-form $\omega$ on an n-dimensional differentiable manifold M, find all integral submanifolds of the exterior differential system that it defines:*

$$\omega = 0.$$

> *In particular, find the maximum dimension of these submanifolds.*

Now, if the coordinate system $(U, x^i)$ is *adapted* to the submanifold $x$ then the coordinate systems will take the form $(u^1, \dots, u^p, x^{p+1}, \dots, x^n)$. Thus, for an integral submanifold one will have that $x^{p+1}, \dots, x^n$ are all constants, which would make:

$$dx^r = 0 \qquad (r = p+1, \dots, n) \tag{1.10}$$

in that case. Thus, since $du^a$ are not generally zero, in order for $\omega$ to be zero, one would have to have $\omega_a = 0$, $a = 1, \dots, p$, and $\omega$ would reduce to the form:

---

([1])  A good historical cross-section of books on exterior differential systems is given by Slebodzinski [**13**], Choquet-Bruhat [**14**], and Bryant, Chern, et al. [**15**].

$$\omega = \sum_{r=p+1}^{n} \omega_r dx^r .\tag{1.11}$$

If $p$ were maximal then $n - p$ would be minimal, and one would then find that $\omega$ could be expressed in terms of either of two *canonical (or normal) forms:*

$$\omega = \sum_{h=0}^{n-p-1} \mu_h dv^h \qquad \text{or} \qquad \omega = dv^0 + \sum_{h=1}^{n-p-1} \mu_h dv^h .\tag{1.12}$$

(This fact was already known to Frobenius [**6**].)

Actually, although a Pfaffian form can have two possible normal forms, nonetheless, a Pfaffian *equation* can only have one, since if one of the component functions – say, $\mu_0$ – is everywhere non-zero then one can divide the components of the first expression in (1.12) by that component and obtain an expression of the second form, which will usually be expressed in the canonical form:

$$\omega = d\lambda - \mu_r dv^r .\tag{1.13}$$

As long as one is setting $\omega$ equal to zero, the division by a non-zero component will not change the nature of the integral submanifolds. For example, whether $\omega = d\lambda$ or $\omega = \mu$ $d\lambda$, the maximal integral submanifolds of $\omega = 0$ are the level hypersurfaces of $\lambda$.

One should note, however, that the process of dividing the component covector ($\mu_0$, …, $\mu_{n-p-1}$) by one of the non-zero components is also how one gets from the homogeneous coordinates of $\mathbb{R}P^n$ to its inhomogeneous coordinates, and since $\omega$ is really defined only up to multiplication by a non-zero function on $M$, as long as one always sets $\omega$ to zero, that one is really concerned with elements $[\omega]$ of the *projectivized cotangent bundle $PT^*M$*, rather than the cotangent bundle $T^*M$ itself. Indeed, in the language of projective analytic geometry [**16**], one can also say that the integral submanifolds $x$ are the *loci* of the points parameterized by the $u$ or the *envelopes* of the families of tangent hyperplanes that are defined by the annihilating hyperplanes of $\omega$.

Since the coordinates $v^h$ are normal to the integral submanifolds, one sees that finding the degree of integrability of $\omega$ is equivalent to finding a normal form for it. However, the process of finding a normal form has a more algebraic character to it than the process of integrating systems of partial differential equations, as we will see in the next section.

## 2. The integrability of the Pfaff problem.

**2. The integrability of the Pfaff problem.** The picture that eventually emerged for the Pfaff problem was that the degree of integrability of the Pfaff problem was indexed by a series of exterior differential forms of increasing degree that we shall denote by $\mathcal{I}_p$, namely:

$$\mathcal{I}_0 = \omega, \qquad \mathcal{I}_1 = d\omega, \qquad \mathcal{I}_2 = \omega \wedge d\omega, \quad \mathcal{I}_3 = d\omega \wedge d\omega, \ldots$$

One sees that there are two basic recursions going on here: In one case, one applies the operator $d$ of exterior differentiation to forms of odd degree, and in the other case, one applies the operator $e_\omega = \omega \wedge$ of exterior multiplication by $\omega$ to forms of even degree. In general, one then has an odd series and an even series:

$$\mathcal{I}_p \in \Lambda^{p+1}\mathrm{M} = \begin{cases} \mathcal{I}_{2k-1} = d\mathcal{I}_{2k-1} = (\mathcal{I}_1)^{\wedge k}, \\ \mathcal{I}_{2k} = \mathcal{I}_0 \wedge \mathcal{I}_{2k-1}. \end{cases} \tag{2.1}$$

We shall introduce the terminology that the $\mathcal{I}_p$ are the successive *integrability forms of degree $p$* for $\omega$, rather than Cartan's terminology [**10**] that made them the successive "derivatives" of $\omega$, since there is really only one differentiation of $\omega$ involved.

In the second edition of his *Ausdehnunglehre* [**5**], Grassmann introduced the notion that the *class* of the 1-form $\omega$ was the minimum number $p$ of independent functions $\mu_h$, $v^h$ that were necessary in order to express $\omega$ in its normal form.

$$\omega = \sum_{h=1}^{m} \mu_h(v)\, dv^h \qquad \text{or} \qquad \omega = d\lambda - \sum_{h=1}^{m-1} \mu_h(v)\, dv^h. \tag{2.2}$$

Hence, depending upon the normal form for $\omega$, the class of $\omega$ would either be $2m$ or $2m - 1$, respectively. Since the equations $v^h = $ const. ($\lambda = $ const., $v^h = $ const, resp.) parameterize the maximal integral submanifolds, and there are $m$ of them, one then sees that the class $p$ of $\omega$ is directly related to the codimension $m$ of the maximal integral submanifolds by way of:

$$p = \begin{cases} 2m \\ 2m-1 \end{cases}, \tag{2.3}$$

depending upon the normal form.

It was Cartan [**10**] who first showed that the class of a Pfaffian form $\omega$ is equal to the smallest positive integer $p$ such that $\mathcal{I}_p$ vanishes. Thus, one has a purely algebraic algorithm for establishing the degree of integrability of a Pfaffian form.

When one considers the sequence $\mathcal{I}_0$, $\mathcal{I}_1$, …, one must recall that, ultimately, one must always recall that when an exterior form has a degree that exceeds the dimension of the manifold, it must vanish from dimensional considerations. Thus, for two-dimensional manifolds, only $\mathcal{I}_0$ and $\mathcal{I}_1$ can be non-zero, so a non-zero Pfaff form can take on only two possible canonical forms: $d\lambda$ and $\mu\, dv$. The only possible integral submanifolds are then curves.

For three-dimensional manifolds, only $\mathcal{I}_0$, $\mathcal{I}_1$, $\mathcal{I}_2$ can be non-zero, so a non-zero Pfaff form can take on only three possible canonical forms, namely:

$$d\lambda, \qquad \mu\, dv, \quad d\lambda + \mu\, dv,$$

which correspond to $\mathcal{I}_1 = 0$, $\mathcal{I}_2 = 0$, $\mathcal{I}_3 = 0$ as the first vanishing term in the series, resp. The maximal integral manifolds in each case have dimensions 2, 2, and 1, resp.

Similarly, for four-dimensional manifolds, one has the same three possibilities that exist on three-manifolds, but also the fourth possible canonical form $\mu_1 \, dv^1 + \mu_1 \, dv^1$, which corresponds to the possibility that $\mathcal{I}_3 = d\omega \wedge d\omega$ is non-vanishing, but $\mathcal{I}_4 = \omega \wedge d\omega \wedge d\omega$ vanishes. The maximal integral manifolds in the four-dimensional case can also be three-dimensional, as well as having dimension one or two, depending upon the degree of integrability.

Some of the terms in the sequence of $\mathcal{I}_p$ deserve special attention since they were introduced independently of the general picture that eventually evolved.

As we know by now, the vanishing of $\mathcal{I}_1 = d\omega$ is locally equivalent to the possibility that there is some local 0-form $v$ such that:

$$\omega = dv. \tag{2.4}$$

Of course, one sees that this is usually only true locally, since when $M$ is multiply connected there will be closed 1-forms that are not exact.

Exactness is what one usually expects of "complete" integrability, in the sense of the existence of integral submanifolds of maximum dimension – viz., one of codimension $p = 1$. In coordinates and components, equation (2.4) is then equivalent to the following system of $n$ first-order partial differential equations for $v$:

$$\frac{\partial v}{\partial x^i} = \omega_i(x^j). \tag{2.5}$$

However, according to *Frobenius's theorem* [**6, 17**] the existence of maximal integral submanifolds is equivalent to the vanishing of the next term in the sequence, namely the 3-form $\mathcal{I}_2 = \omega \wedge d\omega$, which we then refer to as the *Frobenius 3-form* that is associated with the *Pfaff form* $\omega$. The Frobenius 3-form has the local expression:

$$\omega \wedge d\omega = -\frac{1}{3!} \, [\omega_i(\omega_{j,k} - \omega_{k,i}) + \omega_j(\omega_{k,i} - \omega_{i,k}) + \omega_k(\omega_{i,j} - \omega_{j,i})] \, dx^i \wedge dx^j \wedge dx^k, \tag{2.6}$$

so the vanishing of this 3-form is locally equivalent to the first-order partial differential equation for the $\omega_i$ (the other cyclic permutations of $ijk$ give identical equations)

$$\omega_i \left( \frac{\partial \omega_j}{\partial x^k} - \frac{\partial \omega_j}{\partial x^k} \right) + \omega_j \left( \frac{\partial \omega_k}{\partial x^i} - \frac{\partial \omega_i}{\partial x^k} \right) + \omega_k \left( \frac{\partial \omega_i}{\partial x^j} - \frac{\partial \omega_j}{\partial x^i} \right) = 0, \tag{2.7}$$

which is the form that one commonly encounters in the classical literature when one sets $ijk = 123$.

Although the vanishing of $\omega \wedge d\omega$ is not apparently the same thing as the vanishing of $d\omega$, nonetheless, they both lead to the existence of integral hypersurfaces. The vanishing of $\omega \wedge d\omega$ amounts to the existence of an *integrating factor* for the Pfaff form $\omega$, which says that one can have:

$$\omega = \mu\, dv \tag{2.8}$$

without having $\omega = dv$ for any $v$, and still satisfy the Pfaff equation, since the vanishing of $\omega$ is equivalent to the vanishing of $dv$ as long as $\mu$ is non-zero on $M$. In particular, one must have:

$$d\omega = d\mu \wedge dv. \tag{2.9}$$

In the case where $\omega$ has zeroes on $M$, one would say that $\omega$ has *singularities* at those zeroes. Although the zeros of the 1-form $\omega$ are presumed to be dual to the zeroes of some vector field $X$ in some way, nonetheless, there is a significant difference between the character of those zeroes in terms of the solutions to the differential equations that they define. A zero of a vector field is a fixed point of the local flow for the system of ordinary differential equations that are defined by making $X$ the velocity vector field for some congruence of differentiable curves in $M$. By Frobenius's theorem, although global *flows* do not have to exist on $M$ (which implies a common parameter to all integral curves), nevertheless, a global *foliation* of $M$ of dimension one always exist. By comparison, codimension-one foliations *do not* always have to exist, as one sees from the Frobenius condition. Thus, the structure of the singularities – i.e., zeroes – of 1-forms is generally more complicated than the study of the singularities of vector fields (see, e.g., [**18**]). However, they are locally topologically equivalent as far as the *Poincaré-Hopf theorem* ([1]) is concerned, since the map that takes tangent vector to cotangent vectors is presumed to be a bundle isomorphism.

The 4-form $\mathcal{I}_3 = d\omega \wedge d\omega$ deserves special attention because it relates to the fact that when it does not vanish one cannot express $d\omega$ in the form of an exterior product of two exact 1-forms; i.e., $d\omega$ would not be *decomposable*. However, one finds that $d\omega$ can be put into the form:

$$d\omega = d\mu_1 \wedge dv^1 + d\mu_2 \wedge dv^2 = d(\mu_1\, dv^1 + \mu_2\, dv^2), \tag{2.10}$$

The number of independent functions that are required to express $\omega$ is now four, instead of two, which is also the minimum number of 1-forms that are required to express the 2-form $d\omega$. This minimum number of 1-forms that are required in order to express any $k$-form is called its *rank*. It also equals the first vanishing exterior power of the $k$-form. One also sees from (2.10) that if the class of $\omega$ is $2m$ or $2m-1$ then the rank of $d\omega$ is $2m$, in either case. Thus, $\mu\, dv$ and $d\lambda + \mu\, dv$ both give $d\omega$ of rank two, while $\mu_1\, dv^1 + \mu_2\, dv^2$ gives a $d\omega$ of rank four.

---

([1]) Recall that the Poincaré-Hopf theorem says that a non-zero vector field can exist on $M$ iff the Euler-Poincaré characteristic of $M$ vanishes. We shall have no further use for this fact in the present article, so we simply refer the curious to the literature on the subject [**19**].

We summarize some of the elementary cases of $\omega$ in the form of Table 1. We have gone only as far as class 4, since that is all that is necessary for manifold of dimension not greater than four; however, higher dimensional manifolds can occur in thermodynamics.

Table 1. Pfaffian forms of increasing class.

| Class $p$ | Codimension $m$ | Normal form $\omega =$ | $d\omega$ | First vanishing $\mathcal{I}_p$ |
|---|---|---|---|---|
| 1 | 1 | $d\lambda$ | 0 | $d\omega$ |
| 2 | 1 | $\mu\, dv$ | $d\mu \wedge dv$ | $\omega \wedge d\omega$ |
| 3 | 2 | $d\lambda + \mu\, dv$ | $d\mu \wedge dv$ | $d\omega \wedge d\omega$ |
| 4 | 2 | $\mu_1\, dv_1 + \mu_2\, dv_2$ | $d\mu_1 \wedge dv_1 + d\mu_2 \wedge dv_2$ | $\omega \wedge d\omega \wedge d\omega$ |

**3. Symplectic structures.** It gradually emerged from the geometry and mechanics of the early Twentieth Century that certain recurring themes always asserted themselves in Hamiltonian mechanics and the Hamilton-Jacobi approach to wave mechanics. In particular, certain 1-forms seemed to play a fundamental role, namely, the canonical 1-form $\theta$ on the cotangent bundle $T^*M$ for Hamiltonian point mechanics and the contact 1-form on the manifold $J^1(M; \mathbb{R})$ of 1-jets of differentiable functions on a manifold $M$ for wave mechanics. In this section, we shall discuss the former 1-form, while in the following section we shall discuss the latter one.

If $M$ is a differentiable manifold of dimension $n$ whose local coordinate charts take the form $(U, x^i)$ then its cotangent bundle $\pi: T^*M \rightarrow M$ is a differentiable vector bundle whose total space $T^*M$ is a differentiable manifold of dimension $2n$ with coordinate charts that take the form $(U \times \mathbb{R}^{n*}, (x^i, p_i))$, in which the $n$ functions $p_i$ are coordinates on the fiber $T_x^*M$. By definition, the elements of $T_x^*M$ are linear functions on the tangent vectors of $T_xM$. Physically, one can think of covectors as representing either kinematical variables, such as covelocity and frequency-wave number, or dynamical quantities, such as momentum, which is customary for the purposes of Hamiltonian point mechanics. Hence, although one usually sees $\theta$ expressed in terms of components that suggest momentum, nonetheless, that is not the only possible physical interpretation.

The canonical 1-form $\theta$ on $T^*M$ can be described by the somewhat esoteric device of saying that at each point $p_x \in T_x^*M$ the covector $\theta_{(x,p)} \in T_p^*T_x^*M$ has the property that if $X \in T_pT_x^*M$ and $\pi_* X \in T_xM$ is its push-forward by the projection $\pi$ then one must have:

$$\theta_{(x,p)}(X) = p_x(\pi_* X) \tag{3.1}$$

for each $x \in M$ and each $p_x \in T_x^* M$ .

One can also characterize $\theta$ by the fact that for any section $p$: $M \rightarrow T^*M$ one will have:

$$p^* \theta = p. \tag{3.2}$$

which is one way of seeing what makes $\theta$ "canonical."

The definition of $\theta$ becomes more useful for the purposes of physics when one looks at its form in local coordinates $(x^i, p_i)$ for an open set $U \times \mathbb{R}^{n^*} \subset T^*M$. It then takes the local form:

$$\theta = p_i \, dx^i, \tag{3.3}$$

in which it must be stressed that the $p_i$ are coordinate functions on $U \times \mathbb{R}^{n^*}$, not functions on $U$. In fact, the latter situation emerges only when one looks at a section of the cotangent bundle over $U$; i.e., a local 1-form on $U$. If we call that section $p$: $U \rightarrow T^*M$ then one sees that:

$$p^* \theta = p_i(x) \, dx^i = p|_U , \tag{3.4}$$

which is what we expected from (3.2).

Let us treat this canonical 1-form $\theta$ as we have treated every previous Pfaff form, which then defines the predictable Pfaff equation by way of $\theta = 0$. The algebraic solution of this equation is a hyperplane in each $T_p T_x^* M$ , which will then be a vector space of dimension $2n - 1$. We immediately see that since the tangent vectors to $T^*M$ will have the local form $X = X^i \dfrac{\partial}{\partial x^i} + X_i \dfrac{\partial}{\partial p_i}$ , the ones that are annihilated by $\theta$ will then satisfy:

$$p_i X^i = 0, \qquad\qquad X_i = \text{arbitrary}.$$

We then ask the usual question of what the degree of integrability is for this Pfaff equation, and thus examine the sequence $\theta$, $d\theta$, $\theta \wedge d\theta$, …, which becomes:

$$p_i \, dx^i, \qquad\qquad dp_i \wedge dx^i, \qquad\quad - p_i \, dp_j \wedge dx^i \wedge dx^j, \qquad …$$

Since the set of $2n$ 1-forms $\{dx_i, dp^i, i = 1, …, n\}$ are linearly independent, the local $2n$-form on $T^*M$ that they define by taking the exterior product of all of them is non-zero, and, in fact, it is a unit-volume element on $T^*M$ that is usually referred to as the *Liouville volume element*. If we set $\Omega = d\theta$ then it can be expressed in the form:

$$V_{\mathcal{L}} = \Omega \wedge … \wedge \Omega. \text{ ($n$ factors).} \tag{3.5}$$

The fact that the exterior product of $n$ factors is non-vanishing, while the next exterior power of $\Omega$ must vanish, expresses the idea that the 2-form $\Omega$ has *rank n*. As a result, the maximum dimension for integral submanifolds of $\theta = 0$ is $n$, while the Pfaff form $\theta$ is already in its canonical form, namely, $p_i \, dx^i$. We will see shortly why such an integral submanifold is called a "Lagrangian" submanifold.

One can also look at the integrability of the derived system $\Omega = d\theta = 0$. Although the annihilating subspace of the 2-form $\Omega$ in each $T_p T_x^* M$ will be of codimension two, now – i.e., its dimension will be $2(n-1)$ – nonetheless, the maximum integral submanifold will have dimension $n$, as before. If one looks at the local form of the condition that an $n$-dimensional submanifold $f: N \to T^*M$, $u^a \mapsto (x^i(u), p_i(u))$ be an integral submanifold of $\Omega = 0$, which is that $f^*\Omega = 0$, then one finds that:

$$f^*\Omega = f^*(dp_i \wedge dx^i) = \left( \frac{\partial p_i}{\partial u^a} du^a \right) \wedge \left( \frac{\partial x^i}{\partial u^b} du^b \right) = \tfrac{1}{2} [u^a, u^b] \, du^a \wedge du^b, \quad (3.6)$$

in which we have introduced the *Lagrange bracket:*

$$[u^a, u^b] = \frac{\partial p_i}{\partial u^a} \frac{\partial x^i}{\partial u^b} - \frac{\partial p_i}{\partial u^b} \frac{\partial x^i}{\partial u^a}. \qquad (3.7)$$

Thus, $f$ is an integral submanifold iff:

$$[u^a, u^b] = 0. \qquad (3.8)$$

This would seem to justify the popular modern terminology of calling maximal integral submanifolds of $\Omega = 0$ *Lagrangian submanifolds* of $T^*M$.

In particular, a section $p: M \to T^*M$, takes the local form $x^a \mapsto (x^i, p_i(x))$, so the $u$ parameters are the $x$ coordinates in this case. Hence:

$$[u^a, u^b] = \frac{\partial p_i}{\partial x^a} \frac{\partial x^i}{\partial x^b} - \frac{\partial p_i}{\partial x^b} \frac{\partial x^i}{\partial x^a} = \frac{\partial p_b}{\partial x^a} - \frac{\partial p_a}{\partial x^b}, \qquad (3.9)$$

so:

$$p^*\Omega = dp, \qquad (3.10)$$

and a section of the cotangent bundle – i.e., any 1-form – defines a Lagrangian submanifold of that bundle iff it is closed. One could also verify this by noting that since pull-backs commute with exterior differentiation, one will have:

$$p^*\Omega = p^*d\theta = d(p^*\theta) = dp. \qquad (3.11)$$

If we interpret the 1-form $p$ as either the covelocity 1-form or the momentum 1-form for an extended mass distribution then we see that the integral submanifolds will have to represent irrotational motion. Hence, one way of characterizing non-integrability is to say that generally one is considering vorticial motion, which we shall discuss in the second part of this paper.

Now, the 2-form $\Omega = d\theta$ is clearly closed and can be shown to satisfy the non-degeneracy condition that the linear map $i_\Omega : T(T^*M) \to T^*(T^*M)$, $X \mapsto i_X\Omega$ is a linear isomorphism. Thus, it defines what is commonly called a *symplectic structure* on the manifold $T^*M$. Most of the advantages of introducing such a structure center around the fact that a symplectic structure appears naturally in the context of Hamilton's equations of motion for conservative mechanical systems that involve only perfect holonomic constraints. In particular, the formulation of Hamilton's equations, the notion of canonical transformations, and the introduction of Poisson brackets are all quite naturally defined in terms of a symplectic structure.

However, as Lie observed [**20**] as early as 1889, all of the same things can be derived from the nature of contact transformations, which relate to $\theta$, more than $\Omega$. Therefore, since we are more concerned with the specific problem of integrating Pfaff equations at the moment, we shall merely direct the curious to the usual literature on symplectic mechanics [**21-25**].

**4. Contact structures.** The symplectic structure on the cotangent bundle is a special case of a more general situation that takes the form of *contact geometry*. Loosely speaking, a *contact element* at a point in a space $M$ – which we assume to be a differentiable manifold – is a linear subspace in its tangent space. For instance, the line that is tangent to a differentiable curve at a point is a one-dimensional contact element, although a line that intersects the curve transversely would be a "non-integrable" example of a contact element. At the next dimension of contact, one might consider the "osculating plane" (if there is one) to that same point, which is defined by the limit of the plane spanned by the tangent line and a line through the chosen point and any distinct neighboring point as that point approaches the chosen point. Thus, it is also spanned by the velocity and acceleration vectors, so when the acceleration of the curve is collinear with the velocity there will be no osculating plane.

Since one is looking at lines, planes, and $k$-planes more generally, one sees that the geometry of contact elements has much to do with projective geometry. However, the most natural way of introducing contact elements into tangent spaces is by way of "jets of differentiable maps." For instance, let $\Pi$ be a $p$-dimensional parameter manifold, which might simply be an open subset of $\mathbb{R}^p$, and let $M$ be an $m$-dimensional manifold. If $u = (u^1, \ldots, u^p)$ is a point in $\Pi$ and $x \in M$ is a point in $M$ then a *1-jet at $u$* is an equivalence class of differentiable maps $f : U \to M$ such that $U$ is some neighborhood of $u$ and $f(u) = x$, and one further has that $df|_u = df'|_u$ for any two maps $f$, $f'$ in the equivalence class, which we denote by $j_u^1 x$. If one chooses one frame for $T_u\Pi$ and another for $T_xM$, such as the natural frames $\partial_a$ and $\partial_i$ that are defined by a coordinate system $(U, u^a)$ in the neighborhood of $u$ and another one $(V, x^i)$ in the neighborhood of $x$, then the differential maps $df|_u$ in a 1-jet $j_u^1 x$ will all have the same matrix $x_a^i$. One can then uniquely define the 1-jet $j_u^1 x$ by the coordinates $(u^a, x^i, x_a^i)$, which also define local coordinate systems on the set $J^1(\Pi, M)$ that we call the *manifold of 1-jets of differential maps from $\Pi$ to $M$* ([1]).

_______________

([1]) For a treatment of jet manifolds that is tailored to the concerns of mechanics, one can confer Gallisot [**21**], and for a more modern mathematical treatment, one can confer Saunders [**26**].

The way that this relates to contact elements in tangent spaces is shown by recalling that since the map $df|_u : T_u\Pi \to T_xM$ is linear, its image is a linear subspace of $T_xM$ whose dimension equals the rank of $df|_u$; thus, it is a contact element at $x$.

Conversely, a $k$-dimensional linear subspace of $T_xM$ can be associated with a linear map from $\mathbb{R}^k$ to $T_xM$ that has maximal rank – i.e., a linear injection. Such a map also defines a $k$-*frame* in $T_xM$ by the images of the canonical frame vectors on $\mathbb{R}^k$. Thus, one can associate any $k$-frame in $T_xM$ with the 1-jet of a differentiable map from $0 \in \mathbb{R}^k$ to $x \in M$ whose differential has maximal rank $k$; i.e., the map must be a local *immersion* about 0.

In order to relate this picture to the cotangent bundle of $M$ one needs only to consider the manifold $J^1(M; \mathbb{R})$ of 1-jets of differentiable real functions on $M$. It will have coordinate charts that look like $(x^i, f, p_i)$ and, in fact, the *target projection* $\beta: J^1(M; \mathbb{R}) \to \mathbb{R}$, $j_x^1 f \mapsto f$ has fibers at each $f \in \mathbb{R}$ that look like $T^*M$. Dually, if one considers the manifold $J^1(\mathbb{R}; M)$ of 1-jets $j_t^1 x$ of differentiable curves in $M$ then the fibers of the *source projection* $\alpha: J^1(\mathbb{R}; M) \to \mathbb{R}$, $j_t^1 x \mapsto t$ look like the tangent bundle $T(M)$.

Thus, within the purview of jet manifolds one finds that one can be dealing with tangent and cotangent objects of all descriptions – e.g., vectors, covectors, linear subspaces, frames, coframes – in a single natural framework that involves only elementary calculus at its most elementary level.

In order to relate jet manifolds to Pfaffian equations and their integrability, one must first note that there is nothing to say, *a priori*, that a given matrix $x_a^i$ will represent the Jacobian matrix of some set of differentiable functions; for one thing, one would have to make $x_a^i$ the value taken by some differentiable function. This begs the question "On what?" whose answer admits three canonical possibilities: For the general case of $J^1(\Pi, M)$, we have already described the source and target projections, and there is a third one that we shall call the *contact projection*, namely, $\pi_0^1: J^1(\Pi, M) \to \Pi \times M$, $j_u^1 x \mapsto (u, x)$.

The various ways of making $x_a^i$ into a differentiable function that immediately come to mind are the sections of these projections.

A section of the source projection $s: \Pi \to J^1(\Pi, M)$, $u \mapsto s(u)$ will take the local coordinate form:

$$s(u) = (u^a, x^i(u), x_a^i(u)). \tag{4.1}$$

The question of integrability then takes the form of asking whether there exists a differentiable function $x: U \subset \Pi \to M$, $u \mapsto x(u)$ such that $j^1 x(u) = s(u)$ for every $u \in U$, in which we defined the *1-jet prolongation of $x$*, which we denote by $j^1 x$. It then has the local coordinate form:

$$j^1 x(u) = (u^a, x^i(u), x_{,a}^i(u)), \tag{4.2}$$

in which the comma implies partial differentiation with respect to $u^a$ A section $s$ of the source projection is then defined to be *integrable* iff it is the 1-jet prolongation of some differential function $x: \Pi \to M$ – i.e., $s = j^1 x$ – which leads to the local condition:

$$x_a^i = \frac{\partial x^i}{\partial u^a}, \tag{4.3}$$

which does, in fact, take the form of an integrability condition.

One can define an operator $D: J^1(\Pi, M) \to \Lambda^1(\Pi) \otimes T(M)$, $s \mapsto Ds$, in which $J^1(\Pi, M)$ really refers to *sections* of the source projection, $\Lambda^1(\Pi)$ represents the 1-forms on $\Pi$, and $T(M)$ represents vector fields on $M$. Although one can define the operator schematically by $Ds = s - j^1 x$, it is easier to understand in its local expression:

$$Ds(u^a, \ x^i(u), \ x_a^i(u)) = \left( x_a^i(u) - \frac{\partial x^i}{\partial u^a} \right) du^a \otimes \frac{\partial}{\partial x^i} \,. \tag{4.4}$$

One then sees, from (4.3) that a section $s$ is integrable iff $Ds = 0$. This operator $D$ is referred to as the *Spencer operator*, and is the basis for a much more involved treatment of the formal integrability of overdetermined systems of partial differential equations.

One finds that the manifold $J^1(\Pi, M)$ admits a set of $m$ 1-forms $\Theta^i$, which are easiest to describe locally, namely:

$$\Theta^i = dx^i - x_a^i \, du^a. \tag{4.5}$$

When one pulls down these 1-forms to 1-forms on $\Pi$ by way of a section $s$ of the source projection, one obtains:

$$s^* \Theta^i = \left( \frac{\partial x^i}{\partial u^a} - x_a^i \right) du^a, \tag{4.6}$$

which, from (4.3), all vanish iff $s$ is an integrable section.

One can then see that these 1-forms relate very intimately to the Spencer operator by the fact that from (4.6) and (4.4), one gets:

$$Ds = s^* \Theta^i \otimes \frac{\partial}{\partial x^i} \,. \tag{4.7}$$

Thus, the vanishing of $Ds$ is clearly equivalent to the vanishing of $s^* \Theta^i$ for all $i$.

A section of the contact projection will be called a *field of contact elements.* In the context of geodesic fields, one refers to such sections as *slope fields.*

One can also treat the system of $m$ exterior differential equations:

$$\Theta^i = 0, i = 1, \ldots, m \tag{4.8}$$

as a Pfaffian system.

The algebraic solution to this system at each point of $J^1(\Pi, M)$ is a linear tangent subspace of dimension $(p + m + pm) - m = p(1 + m)$, which is then the maximum possible dimension for an integral submanifold, at least naively. The actual detailed analysis of the integrability of this system can be quite involved, since it is basically a system of first-order partial differential equations, so we merely specialize it to the previous context of the manifold $J^1(M; \mathbb{R})$, which then has a single canonical 1-form:

$$\Theta = df - p_i \, dx^i. \tag{4.9}$$

A section of the source projection is a map $s\colon M \to J^1(M; \mathbb{R})$ that locally looks like:

$$s(x) = (x^i, f(x), p_i(x)), \tag{4.10}$$

and which is integrable iff:

$$p_i = \frac{\partial f}{\partial x^i}; \tag{4.11}$$

i.e., $p_i \, dx^i = df$ is an exact form.

The algebraic solution to the exterior system $\Theta = 0$ is a hypersurface in each tangent space to $J^1(M; \mathbb{R})$, so it will have a dimension of $2m$, since the dimension of $J^1(M; \mathbb{R})$ is $2m + 1$. However, the maximum dimension for an integral submanifold of $J^1(M; \mathbb{R})$ for this Pfaff system is actually $m$, and one calls such a maximal integral submanifold of $J^1(M; \mathbb{R})$ a *Legendrian submanifold*. They are closely related to the Lagrangian submanifolds of $T^*M$ by the fact that, first of all:

$$d\Theta = - dp_i \wedge dx^i = - \Omega. \tag{4.12}$$

(Recall that $T^*M$ is the fiber over each point of $\mathbb{R}$ for the target projection.) Therefore, if $\rho\colon M' \to J^1(M; \mathbb{R})$, $u \mapsto \rho(u)$ is an $m$-dimensional submanifold of $J^1(M; \mathbb{R})$ then:

$$\rho^* d\Theta = - \rho^* \Omega. \tag{4.13}$$

Thus, we see that the integrability situation for integral submanifolds is quite similar to the situation for $T^*M$ "plus one."

A particular example of an $m$-dimensional manifold of $J^1(M; \mathbb{R})$ is given by a section of the source projection, for which $M' = M$. The situations in which one needs to go beyond that class of $m$-dimensional submanifolds of $J^1(M; \mathbb{R})$ are usually the situations in which the target projection of the submanifold has singularities, such as folds, cusps, and the like.

The closed 2-form $d\Theta = -\Omega$ does not define an actual symplectic structure on the tangent spaces to $J^1(M; \mathbb{R})$, which are always odd-dimensional, and therefore cannot admit a non-degenerate 2-form. However, its restriction to the hyperplanes that are "vertical" with respect to the target projection $\beta: J^1(M; \mathbb{R}) \to \mathbb{R}$ does define a symplectic structure. The vertical tangent vectors to that projection project to zero under $d\beta$, so locally a vertical tangent vector to the target projection will look like:

$$V = V^i \frac{\partial}{\partial x^i} + V_i \frac{\partial}{\partial p_i}, \qquad (4.14)$$

which resemble vector fields on $T^*M$, and the 1-forms that are dual to these spaces will then take the local form of 1-forms on $T^*M$.

Just as the symplectic structure that is defined on $T^*M$ by $\Omega$ allows one to define Hamiltonian vector fields for given differentiable function, which locally define systems of first-order ordinary differential equations that take the form of Hamilton's equations, similarly, a differentiable function on $J^1(M; \mathbb{R})$ is associated with a *characteristic vector field*. Since a function on $J^1(M; \mathbb{R})$ locally takes the form $F(x^i, f, p_i)$, one then sees that for an integrable section of the source projection, it takes the form:

$$F(j^1 f) = F\left(x^i, f(x), \frac{\partial f}{\partial x^i}(x)\right), \qquad (4.15)$$

and the level hypersurfaces of the function $F$ then become first-order partial differential equations, for the function $f$ on $M$. The characteristic vector field for $F$ then allows one to solve the initial-value problem for such an equation by Cauchy's method of characteristics. Thus, one finds that the methods of jet manifolds are just as natural to the theory of differential equations as they are to mechanics, field theories, and the calculus of variations.

For more details on Lie's fundamental work on contact transformations, see [**27-29**]. For a more modern application of the methods of contact geometry to wave mechanics, see Arnol'd [**30**].

The fact that 1-forms play a broad-ranging role in physical models might be basically attributed to the fact that generally forces take that form. This then translates into the idea that the fundamental fields of classical physics are basically defined in terms of forces per unit mass, charge, or what have you, at least in the "test particle" approximation. In the case of thermodynamics, one is implicitly discussing a force 1-form when one refers to the 1-form that takes the form of the "differential increment of work done on the system," which is not, in general an exact differential.

If we define a 1-form to be integrable iff it is exact then we see that there are two broad classes of non-integrable 1-forms: If the manifold that it is defined on is multiply-connected then it might be closed, but not exact, or, more generally, it might not be closed, either. One then sees that the discussion of the previous section then allows one to refine the last possibility even further by examining the degree of the integrability of the 1-form in question – i.e., its canonical form.

Actually, such considerations have been implicitly applied to many of the mathematical models of theoretical physics for some time already. However, in most cases, the fact that one was simply applying the general theory of Pfaff forms was not always stated explicitly. Thus, we shall review some of the many physical models in which the degree of integrability of a Pfaff form must be addressed, by starting with the highest degree of integrability – viz., exactness – and then proceeding to successively higher levels.

**5. Force 1-forms.** One of the first times that students of physics are exposed to the difference between exact and inexact differentials is in the context of conservative versus non-conservative forces, and usually by way of the work done by such forces on some physical system. Since that eventually means the conservation or non-conservation of total energy in that system, it is not surprising that the inclusion of non-conservative forces eventually gives way to a more thermodynamic analysis. Upon closer inspection, one often finds that the non-conservation is related to a fundamental incompleteness in the model that makes unmodeled sources or sinks of energy take the form of an inexact contribution to the work done on a given system.

We shall first define force 1-forms in full generality and then show how some of the common non-conservative examples can be approached using the integrability of Pfaff systems and then go one to a more detailed thermodynamical modeling.

Let the *configuration* of a physical object be described by an embedding $x : \mathcal{O} \to M$, $u \mapsto x(u)$ of some subset $\mathcal{O} \subset \mathbb{R}^p$ of a $p$-dimensional parameter space into an $n$-dimensional differentiable manifold $M$. In particular, the parameter space can contain the time parameter $t$. Indeed, if the object in question is point-like then the parameter space will be only $\mathbb{R}$, the time line.

If $(U, x^i)$ is a local coordinate chart on $M$ that contains the image of $x$ then one can express the embedding by means of a system of $n$ equations in $p$ independent variables:

$$x^i = x^i(u^a), \qquad (a = 0, \ldots, p-1, i = 1, \ldots, n). \qquad (5.1)$$

The fact that the map $x$ is an embedding implies that it is a diffeomorphism onto its image in $M$, so, in particular, the differential map $dx|_u : \mathbb{R}^p \to T_{x(u)}M$ must have rank $p$ at every point $u \in \mathcal{O}$. One can also say that its matrix $\partial x^i / \partial u^a$ must have rank $p$ at every point of $\mathcal{O}$ and in every coordinate system.

The *kinematical state* of the object will be defined by a section of the source projection $\alpha : J^1(\mathcal{O}; M) \to \mathcal{O}$. It is thus a differentiable map $s : \mathcal{O} \to J^1(\mathcal{O}; M)$, $u \mapsto s(u)$ that projects back to the identity map on $\mathcal{O}$: $\alpha \cdot s = I_{\mathcal{O}}$. In a coordinate chart for $J^1(\mathcal{O}; M)$ that looks like $(u^a, x^i, x^i_a)$, $s(u)$ will then look like $(u^a, x^i(u), x^i_a(u))$; i.e., the system of equations:

$$x^i = x^i(u), \; x^i_a = x^i_a(u). \qquad (5.2)$$

In the simplest case of point mechanics, where $\mathcal{O}$ is simply an interval $[t_0, t_1]$ along the time line, this becomes:

$$x^i = x^i(t), \; x^i_a = x^i_a(t). \qquad (5.3)$$

In that case, we shall also denote the generalized velocities by $v^i$, instead of $x^i_0$.

The kinematical state that is described by $s$ is integrable iff the $x^i_a(u)$ are the generalized velocities that are associated with the $x^i(u)$:

$$x^i_a = \frac{\partial x^i}{\partial u^a}, \qquad (5.4)$$

so for point motion this will take the form:

$$v^i = \frac{dx^i}{dt}. \qquad (5.5)$$

Thus, the vanishing of the Spencer operator $Ds$ has a simple physical interpretation in this case. That is, an integrable section of the source projection of $J^1(\mathcal{O}; M)$ describes a kinematical state in which the velocity vector field is, in fact, the time derivative of the position in $M$. A counter-example of a non-integrable section might be given by a velocity that is described in an anholonomic frame field.

A *virtual displacement* (or *variation*) of a kinematical state $s$ is a vector field $\delta s$ that is defined on the image of $s$, or, as one sometimes says, *along s*. Thus $\delta s : \mathbb{R}^p \to T(x(\mathcal{O}))$, $u \mapsto \delta s(u)$, where $\delta s(u)$ is a vector in $T_{s(u)}J^1(\mathcal{O}; M)$    Its local expression in terms of coordinates will then be:

$$\delta s(u) = \delta u^a(u) \frac{\partial}{\partial u^a} + \delta x^i(u) \frac{\partial}{\partial x^i} + \delta x_a^i(u) \frac{\partial}{\partial x_a^i} .$$

The *dynamical state* of the object $\mathcal{O}$ will be defined by its response to any virtual displacement $\delta s$. This takes the form of a 1-form $\theta$ on $J^1(\mathcal{O}; M)$ that gives a scalar function $\delta W$ on $\mathcal{O}$ that one calls the *virtual work* that is performed by the virtual displacement $\delta s$:

$$\delta W = \theta(\delta s). \tag{5.6}$$

Locally, the 1-form $\theta$ will look like:

$$\theta = P + F + \Pi = P_a \, du^a + F_i \, dx^i + \Pi_i^a dx_a^i . \tag{5.7}$$

Thus, one will also have:

$$\delta W = P(\delta u) + F(\delta x) + \Pi(\delta v) = P_a \, \delta u^a + F_i \, \delta x^i + \Pi_i^a \, \delta x_a^i . \tag{5.8}$$

The physical character of each part of the 1-form $q$ is then clear:

1) $P$ is a 1-form on $\mathcal{O}$ that represents the power that is being applied or dissipated, independently of external forces or its motion.

2) $F$ is a 1-form on $M$ that describes the external forces that are applied at every point of the manifold. Thus, $F$ will not describe localized forces that are applied at isolated points, unless one wishes to include discontinuous components.

3) $\Pi$ represents the generalized momenta that are associated with the dynamics of $\mathcal{O}$.

The functional relationships that define the components of $\theta$:

$$P_a = P_a(u, x, v), \qquad F_a = F_a(u, x, v), \qquad \Pi_i^a = \Pi_i^a(u, x, v) \tag{5.9}$$

then amount to a set of *mechanical constitutive laws* that associate kinematical states with dynamical states.

From now on, we shall focus exclusively on the force part of $\theta$, and also assume that we are dealing with point mechanics. Thus:

$$F = F_i(t, x, v) \, dx^i. \tag{5.10}$$

This makes:

$$dF = \frac{\partial F_i}{\partial t} dt \wedge dx^i - \frac{1}{2}\left( \frac{\partial F_i}{\partial x^j} - \frac{\partial F_j}{\partial x^i} \right) dx^i \wedge dx^j - \frac{\partial F_i}{\partial v^j} dx^i \wedge dv^j . \tag{5.11}$$

Thus, one can already see the possible obstructions to the integrability of $F$, which we will interpret to mean exactness, namely, the possibility that $F$ might be time-varying, it might not admit a force potential, and it might depend upon the velocity of motion.

Of course, the simplest case is that of a time-invariant force that is independent of the velocity and admits a force potential function – $U$ on $M$:

$$F = -dU = -\frac{\partial U}{\partial x^i}\,dx^i. \qquad (5.12)$$

One then sees how there is undeniably incompleteness in such systems that is due to the approximation that went into ignoring time-varying forces, velocity-dependent forces, and even non-conservative forces that depend upon only $x$.

Although the annihilating subspaces Ann$[\theta]$ will be of codimension one, or of dimension $p + n + np - 1$ (which is $2n$ when $p = 1$), the maximal dimension for integral submanifolds will not generally be close to that high in dimension. From the general theory of Pfaffian forms, the maximal dimension will depend upon the nature of $F$, $dF$, and the higher-degree integrability forms. If $\mathcal{I}_{2k}$ or $\mathcal{I}_{2k-1}$ is the first vanishing integrability form then the minimum codimension is $k$, so the maximum dimension of the integral submanifolds will be $p + n + np - k$, which is $2n - k + 1$ when $p = 1$.

Since an integral submanifold $y: \Sigma \to J^1(\mathcal{O}; M)$, $s \mapsto y(s)$, where $\Sigma$ is $k$-dimensional must satisfy:

$$0 = y^*F = \left( F_i\,\frac{\partial y^i}{\partial s^\alpha} \right)ds^\alpha \qquad (\alpha = 1, \ldots, k), \qquad (5.13)$$

one sees that the integral submanifolds must satisfy the local condition:

$$0 = F_i\,\frac{\partial y^i}{\partial s^\alpha} = g(\mathbf{F}, \mathbf{v}_\alpha) \qquad (\alpha = 1, \ldots, k), \qquad (5.14)$$

in which $g$ represents a metric on $T(M)$ that allows one to define a vector field $\mathbf{F}$ on $J^1(\mathcal{O}; M)$ that corresponds to the 1-form $F$, and the vector fields $\mathbf{v}_\alpha$ are the generalized velocities of the images of the parameter curves in $\Sigma$ under $y$:

$$\mathbf{v}_\alpha = \frac{\partial y^i}{\partial s^\alpha}\,\frac{\partial}{\partial y^i}. \qquad (5.15)$$

Thus, the physical interpretation of the components $g(\mathbf{F}, \mathbf{v}_\alpha)$ of $y^*F$ is that they represent generalized power expressions. A submanifold of $J^1(\mathcal{O}; M)$ will then be an integral of $F = 0$ iff all of the vectors tangent to it are orthogonal to the ambient force iff no power is added or dissipated along a curve that is tangent to that vector. Thus, one sees that one dealing with conservation of energy on the integral submanifolds. We shall return to this topic in a later section.

The highest degree of integrability (i.e., integral submanifolds of codimension one) is attained for force 1-forms that look like – $dU$ or $\mu\,dU$, where $\mu$ is a function on $J^1(\mathcal{O}; M)$ that plays the role of an integrating factor. Of course, in many cases, the function $U$ will depend only up the points $x$ of $M$, while being constant with respect to $u$ and $v$. One can then characterize the maximal degree of integrability as being a force that is associated with an energetically-closed physics system; i.e., all degrees of freedom in the flow of energy are accounted for completely.

Thus, in order find examples from physics of force 1-forms that have a lower degree of integrability than co-dimension one, one needs to look at the non-conservation of energy in one form or another.

**6. Non-conservative forces.** Since the non-conservation of energy can generally take the form of energy being lost or gained by a system, we shall give examples both cases. The most common situation in which energy is lost is when there is work done against kinetic friction or viscous drag. The example that we shall give of energy being added to a system comes from electromagnetic induction. One also sees that since a detailed accounting of the force of interaction between sliding solid bodies or solid bodies moving through viscous fluids would be quite intractable, or at least, complicated, the non-conservative nature of such forces – i.e., their non-integrability – can be directly attributed to the incompleteness in the modeling of them.

*a. Kinetic friction/viscous drag.* The essential features of either the force of kinetic friction between solid bodies or the force of viscous drag when the ambient medium is a viscous fluid is that one has a force that always points in the opposite direction to the velocity vector $\mathbf{v}$ with a magnitude that is some function of $t$, $x$, and $v$, if we are to include the possibility that the force is inhomogeneous and time-varying, such as the force of friction on a road during a snowstorm or the aerodynamic drag on wings in turbulent air.

If we define the unit vector in the direction of $\mathbf{v}$ (when it is non-zero) to be $\hat{\mathbf{v}} = \mathbf{v} / \|\mathbf{v}\|$ then we can say that the general form of such forces is either:

$$\mathbf{f}(t, x, v) = -f(t, x, v)\,\hat{\mathbf{v}}\,, \qquad (6.1)$$

or:

$$\mathbf{f}(t, x, v) = -f(t, x, v)\,\mathbf{v}\,, \qquad (6.2)$$

where we have absorbed the factor of $\|\mathbf{v}\|^{-1}$ into the definition of $f$.

Some elementary examples of forces of this form are then:

1. Linear (Coulomb) kinetic friction. One uses the form (6.1) and defines:

$$f = \mu_k\,N, \qquad (6.3)$$

where $\mu_k$ is the coefficient of kinetic friction for the surfaces in contact, and $N$ is the magnitude of the normal force that is applied at the point of contact.

2. Linear viscous drag, which usually applies to low Reynolds number flow regimes. One uses the form (6.2) with:

$$f = \alpha, \qquad (6.4)$$

where $\alpha$ is a constant.

3. Nonlinear viscous drag, which becomes more necessary as the Reynolds number increases. One then uses the form (6.1) with:

$$f = \left(\tfrac{1}{2} C_D \rho A\right) \parallel \mathbf{v} \parallel^2, \tag{6.5}$$

in which $C_D$ is the empirically-determined coefficient of drag for the given body in the given attitude in the given fluid medium in the given flow regime, $\rho$ is the mass density of that medium, and $A$ is the reference cross-sectional area for the body, which is assumed to be normal to $\mathbf{v}$. Of course, there is considerable hidden complexity to the "constant" $C_D$. At the very least, it might change radically as $\parallel \mathbf{v} \parallel$ increases (e.g., boundary-layer separation, transonic flow), as well as when the moving body changes attitude so that the reference cross-section is no longer perpendicular to the motion.

Although velocity vector fields are generally defined only on the objects whose motion is being described, nevertheless, one can define a global 1-form on $J^1(\mathbb{R}; M)$ that represents the force of friction/drag. For the sake of simplicity, though, we shall describe it in terms of its local expression using the coordinates $(t, x^i, v^i)$ for $J^1(\mathbb{R}; M)$, namely:

$$f = -\alpha(t, x, v)\, v_i\, dx^i, \tag{6.6}$$

in which $\alpha(t, x, v)$ is an empirical function that describes the materials in contact. Its form can be deduced from (6.3), (6.4), (6.5), for instance:

$$\alpha(t, x, v) = \begin{cases} \dfrac{\mu_k N}{\parallel v \parallel} & \text{kinetic friction,} \\ \alpha(t, x) & \text{linear viscous drag,} \\ \tfrac{1}{2}(C_D A \rho) \parallel v \parallel & \text{nonlinear viscous drag.} \end{cases} \tag{6.7}$$

One computes:

$$df = -v_i\, d\alpha \wedge dx^i - \alpha\, dv_i \wedge dx^i = -(v_i\, d\alpha + \alpha\, \delta_{ij}\, dv^j) \wedge dx^i. \tag{6.8}$$

Already, one can see that much depends upon the nature of $\alpha$. In the simplest case of linear, homogeneous, time-invariant viscous drag, $\alpha$ is a constant, so all that remains of $df$ is $\alpha\, \delta_{ij}\, dx^i \wedge dv^j$. For a three-dimensional $M$, $\mathcal{I}_4 = df \wedge df \wedge df$ will be the highest non-zero integrability form. We summarize the various possibilities when $\alpha$ is assumed to be time-invariant and homogeneous in the following table, in which we have introduced the unit covector $u_i = v_i / \parallel v \parallel$ in the direction of $v$:

Table 2.  *df* for various types of friction and drag forces.

| Force | $\alpha$ | $d\alpha$ | $df$ |
|---|---|---|---|
| Kinetic friction | $\mu_k \parallel v \parallel^{-1}$ | $-\alpha \parallel v \parallel^{-2} v_i \, dv^i$ | $-\alpha (\delta_{ij} - u_i u_j) \, dv^i \wedge dx^j$ |
| Linear drag | $\alpha$ (constant) | 0 | $-\alpha \, \delta_{ij} \, dv^i \wedge dx^j$ |
| Nonlinear drag | $\frac{1}{2} C_D A \rho \parallel v \parallel$ | $-\alpha \parallel v \parallel^{-2} v_i \, dv^i$ | $-\alpha (\delta_{ij} + u_i u_j) \, dv^i \wedge dx^j$ |

One sees that the coefficients of $dv^i \wedge dx^j$ define a symmetric matrix $f_{ij}$ in any case. In the first case, $\delta_{ij} - u_i u_j$ represents the projection operator onto the plane perpendicular to $v$, while in the second case $\delta_{ij}$ is simply the identity matrix, while in the last case, $\delta_{ij} + u_i u_j$ adds the projection of a vector in the direction of $v$ to the vector. As a result, one gets:

$$\mathcal{I}_5 = df \wedge df \wedge df = -\alpha^3 \det[f_{ij}] \, \mathcal{V}_v \wedge \mathcal{V}_x, \tag{6.9}$$

in which $\mathcal{V}_x$ is the volume element for $x$-space and $\mathcal{V}_v$ is the volume element for $v$-space. Thus, since $\det[f_{ij}]$ vanishes in the first case, one sees that $\mathcal{I}_5$ vanishes in that case, but not the other two, since the eigenvalues are $(1, 1, 1)$ and $(1, 1, 2)$, respectively. However, $5 = 2(3) - 1$, so $k$ is still 3, and the maximum dimension for an integral submanifold in $J^1(\mathbb{R}; M)$ when $M$ is three-dimensional is $1 + 3 + 3 - 3 = 4$. In fact, one finds that the zero section of the contact projection – i.e., $Z(t, x) = (t, x, 0)$ – is such a maximal integral submanifold. Furthermore, any integral curve $\sigma(t) = (t, x(t), v(t))$ for $f = 0$ must satisfy:

$$0 = \sigma^* f = -\alpha (v_i v^i) \, dt = -\alpha \parallel v \parallel^2 dt, \tag{6.10}$$

and since we are only considering non-relativistic motion at the moment, this is true iff $\parallel v \parallel = 0$. That is, an integral curve must represent a point particle at rest.

If one adds a constant (vertical downward) gravitational field, which we describe by an exact 1-form $-mg \, dz = -d(mgz)$, to the force of linear drag then the total forces acting on $m$ will be described by the 1-form:

$$F = -d(mgz) - \alpha v, \tag{6.11}$$

which we define on $J^1(\mathbb{R}; M)$ by way of:

$$f = -\alpha (v_1 \, dx^1 + v_2 \, dx^2) - (mg + \alpha v_3) \, dx^3. \tag{6.12}$$

One immediately sees that $df$ has not changed as a result of this, so $\mathcal{I}_5$ has not, either, and the maximal integral submanifolds are still four-dimensional.

However, the physical interpretation of the maximal integral sections of the contact projection $J^1(\mathbb{R}; M) \to \mathbb{R} \times M$ is somewhat more interesting. $f$ pulls down to essentially itself, except that $\alpha$ and $v_i$ are now functions of $t$ and $x$, so condition for such a section to be integral is that all of the components $f_i$ vanish, which implies that:

$$v_1 = v_2 = 0, \qquad v_3 = -\frac{mg}{\alpha} \equiv -v_T, \tag{6.13}$$

which says that $m$ is falling with a finite terminal velocity $v_T$. This is essentially the "steady-state" solution to the problem of an extended body falling in a linear viscous medium.

*b. Electromagnetic induction.* In electrostatics, if one represents the electric field strength by a 1-form $E = E_i \, dx^i$ on a three-dimensional spatial manifold $\Sigma$ then the fact that one usually demands that $dE = 0$ will imply that (at least, locally) $E$ will take the form $-dU$ for some potential function $U$ on $\Sigma$. Thus, the Pfaffian equation $E = 0$ is completely integrable, and its maximal integral submanifolds are the equipotential surfaces in $\Sigma$.

Now, the constraint that $dE$ must vanish is based in conservation of energy, since one is demanding that the work done on a unit charge by the electric 1-form $E$ around any loop $z_1$, which is (rather incorrectly) referred to as the *electromotive force (e.m.f)*, must vanish. That is:

$$\mathcal{E}[z_1] = \int_{z_1} E = 0. \tag{6.14}$$

However, if $\mathcal{V}_s$ is the volume element on $\Sigma$ then when the loop $z_1$ bounds a 2-chain $c_2$ (i.e., a surface patch) that is linked by a time-varying magnetic field $\#_s \mathbf{B} = i_{\mathbf{B}} \mathcal{V}_s$, which we then regard as a 2-form, Faraday's law of induction makes:

$$\mathcal{E}[z_1] = -\frac{d}{dt} \Phi_{\mathbf{B}}[c_2], \tag{6.15}$$

in which:

$$\Phi_{\mathbf{B}}[c_2] = \int_{c_2} \#_s \mathbf{B}$$

is the total magnetic flux though $c_2$.

Thus, the work done around a loop by the field $E$ is no longer zero, since the process of induction is adding energy to the system.

The differential form of Faraday's law is then obtained by using Stokes's theorem on the left-hand side to first make:

$$\int_{\partial c_2} E = \int_{c_2} d_s E = -\int_{c_2} \#_s \frac{\partial \mathbf{B}}{\partial t}, \tag{6.16}$$

in which $d_s$ refers to the spatial exterior derivative, and then assuming that this is true for all $c_2$:

$$d_s E = - \#_s \frac{\partial \mathbf{B}}{\partial t}. \tag{6.17}$$

Since $d_s E = \#_s$ curl $\mathbf{E}$, one gets the traditional vector calculus form of this:

$$\text{curl } \mathbf{E} = - \frac{\partial \mathbf{B}}{\partial t}.$$

One sees that $E$ cannot be exact, since $\mathcal{I}_1 = d_s E$ is non-vanishing. Hence, we must go to higher-order integrability forms.

One finds that:

$$\mathcal{I}_2 = E \wedge d_s E = -E(\dot{\mathbf{B}}) \, \mathcal{V}_s = - <\mathbf{E}, \, \dot{\mathbf{B}}> \mathcal{V}_s, \tag{6.18}$$

in which  is the spatial volume element. Thus, $\mathcal{I}_2$ will vanish iff $\mathbf{E}$ is perpendicular to $\dot{\mathbf{B}}$.

Since $E$ is spatial, but time-varying, as is $dE$, one will have that:

$$\mathcal{I}_3 = d_s E \wedge d_s E = 0, \tag{6.19}$$

because it will be a 4-form on a three-dimensional space.

If $\mathcal{I}_2$ vanishes first then the minimum codimension for an integral submanifold is still 1, as in the case of an electrostatic potential. However, if $\mathcal{I}_3$ is the first non-vanishing integrability form then the minimum codimension is two. Thus, one cannot define equipotential surfaces in space anymore, but only curves.

Of course, since we are talking about time-varying fields, we really should define them upon a four-dimensional space-time manifold $M$, in which case, equipotential surfaces would be hypersurfaces, while the next lower degree of integrability would involve turning the spatial integral curves into space-time surfaces.

In magnetostatics, one represents the magnetic field strength as a 1-form $B$ on $\Sigma$, and if one introduces an electric current $\mathbf{I}$ as the source of $B$, then the one is coupled to the other by way of:

$$d_s B = \#_s \mathbf{I}. \tag{6.20}$$

Thus, already in the static case, we can see that the 1-form $B$ cannot be exact, except outside the support of $\mathbf{I}$. However, since:

$$\mathcal{I}_2 = B \wedge d_s B = B \wedge \#_s \mathbf{I} = <\mathbf{B}, \mathbf{I}> \mathcal{V}_s, \tag{6.21}$$

one sees that this will vanish, from the right-hand rule. Therefore, the Pfaffian equation is still completely integrable, and thus admits integral surfaces, although one cannot really call them equipotential surfaces, anymore.

When one goes to time-varying electromagnetic fields, Maxwell's law of induction couples the *magnetomotive force (m.m.f.)*:

$$\mathcal{M}[c_1] = \int_{c_1} B \,, \tag{6.22}$$

to the time derivative of the total electric flux that couples a 2-chain $c_2$:

$$\Phi_{\mathbf{E}}[c_2] = \int_{c_2} \#_s \, \mathbf{E} \,, \tag{6.23}$$

by way of:

$$\mathcal{M}[\partial c_2] = \frac{d}{dt} \, \Phi_{\mathbf{E}}[c_2]. \tag{6.24}$$

(Note the positive sign, this time.)

It is also important to note that this coupling is only true outside of the support of any electric current $\mathbf{I}$ that contributes to $B$.

Equation (6.24) takes on the differential form:

$$d_s B = \#_s \frac{\partial \mathbf{E}}{\partial t} \,, \tag{6.25}$$

or

$$\operatorname{curl} \mathbf{B} = \frac{\partial \mathbf{E}}{\partial t} \,. \tag{6.26}$$

Hence, the situation regarding the integral submanifolds for the Pfaffian system $B = 0$ is analogous to the one that we discussed for $E = 0$.

Although one can examine the general case of the exterior differential system $F = 0$ when $F = dt \wedge E + \#_s \mathbf{B}$ is the Minkowski electromagnetic field strength 2-form on the four-dimensional spacetime manifold $M$, nevertheless, since $F$ is a 2-form, not a 1-form, it no longer defines a Pfaffian equation, and thus falls outside the scope of our present discussion. However, we shall return eventually to discuss the integrability of $A = 0$ when $F = dA$.

**7. Linear constraint forces [1, 2].** One way of characterizing holonomic constraints on a mechanical system is to say that they constrain the state – or configuration – of a physical system to some specific subset of the configuration manifold $M$. One can also impose the constraints at the differential level by specifying a subset of the tangent space at each point of $M$. When this subset is a linear subspace that has the same dimension $k$ at every point of $M$ one has a *linear constraint* that also defines a vector sub-bundle $C(M)$ of $T(M)$ that has rank $k$ and which one can call the *constraint bundle* of the physical system. One also refers to the number $k$ as the number of *degrees of freedom* in the system.

Since a vector sub-bundle $C(M)$ of $T(M)$ also defines a differential system, the immediate question is then the integrability of that differential system. When $C(M)$ is completely integrable, all of the $k$-dimensional vector spaces $C_xM$ are tangent to $k$-dimensional submanifolds of $M$ that define the leaves of a foliation of dimension $k$. In particular, the submanifolds have the same dimension as the number of degrees of freedom in the system, so a vector field on $M$ with values in $C(M)$ can be the velocity vector field of a congruence of curves that all lie with the same leaf as they did initially. However, when the constraint bundle $C(M)$ is not completely integrable, one generally finds that the vector fields with values in $C(M)$ can still be integrated into congruences of curves, but they do not have to lie in any specified submanifold of $M$. For instance, an irrational flow on a 2-torus has one degree of freedom, even though the integral curves do not close into circles.

Since linear constraints are usually defined by systems of linear equations, in practice, it is appropriate to assume that the constraint bundle $C(M)$ has fibers that are annihilated by some Pfaffian system $\xi^a = 0$, $a = 1, \ldots, k$. In the case where one has simply one linear constraint $\xi = 0$, one then finds that one can apply the methods of Part I on the integrability of the Pfaffian equation to great advantage. In particular, when $d\xi$ is zero (non-zero, resp.) one calls the constraint *holonomic* (*non-holonomic*, resp.).

In the case of holonomic constraints, since $d\xi = 0$, $\xi$ will (at least, locally) take the form $\xi = d\phi$ for some 0-form $\phi$. Thus, the constraint that $\xi = 0$ is simply the constraint that the motion must be confined to lie in the level surfaces of the function $\phi$. If $d\xi \neq 0$, but $\xi \wedge d\xi = 0$ then one still has complete integrability – i.e., the constraint submanifolds will be of codimension one – but they will not be level surfaces of any function.

A somewhat "canonical" example of a non-holonomic constraint on a mechanical system is defined by rolling without slipping. However, one finds that since the lowest dimension in which one finds non-vanishing 3-forms, such as $\mathcal{I}_2$, is three, one cannot use the example of a tire rolling without slipping along a curve, since the configuration manifold, which has $(x, \theta)$ for its coordinates, is two-dimensional; here, $x$ is the distance along the curve from some reference point and $\theta$ is the angle through which the tire has rotated since some initial reference configuration. Thus, one must, at the very least, consider a tire rolling without slipping on a surface.

If that surface is a plane then the configuration manifold $M$ has coordinates $(x, y, \theta, \psi)$, where $\psi$ is the angle between the plane of the tire and the $xz$-plane, and the rolling-without-slipping constraint says that the translational velocity $\mathbf{v}(t) = (v_x(t), v_y(t))$ of the center of mass or point of contact of the tire with the plane, which is described by the position vector $\mathbf{r}(t) = (x(t), y(t))$ with respect to some chosen origin $O$, must be coupled to the angular velocity $(\omega_T(t), \omega_N(t))$, where $\omega_T(t)$ is the transverse angular velocity (i.e., about the axle) and $\omega_N(t)$ is the angular velocity about the vertical axis, by way of the linear relationship that follows from saying that the velocity is in the plane of the tire and its magnitude is $\omega_T R$, where $R$ is the radius of the tire. Thus if we define the orthonormal 2-frame in the $xy$-plane:

$$\mathbf{e}_L = \cos\psi\, \mathbf{i} + \sin\psi\, \mathbf{j}, \qquad \mathbf{e}_T = -\sin\psi\, \mathbf{i} + \cos\psi\, \mathbf{j}, \qquad (7.1)$$

which is the $\{\mathbf{i}, \mathbf{j}\}$ frame rotated through the angle $\psi$ then we can say that:

$$\mathbf{v} = \omega_{\mathrm{T}} R \, \mathbf{e}_{\mathrm{L}}. \tag{7.2}$$

If one rotates this back to the $\{\mathbf{i}, \mathbf{j}\}$ frame then one finds that:

$$\dot{x} = R \cos \psi \, \dot{\theta}, \qquad \dot{y} = -R \sin \psi \, \dot{\theta}. \tag{7.3}$$

This also says that:

$$\frac{dy}{dx} = -\cot \psi, \tag{7.4}$$

as long as $\dot{\theta}$ is non-vanishing, which means that the tire is indeed rolling.

This suggests that the Pfaffian form that defines the constraint on the velocity vectors is defined by the vanishing of the 1-form:

$$\xi = \sin \psi \, dx + \cos \psi \, dy. \tag{7.5}$$

The fact that the constraint is non-holonomic follows directly from the fact that:

$$\mathcal{I}_1 = d\xi = d\psi \,^\wedge (\cos \psi \, dx - \sin \psi \, dy), \tag{7.6}$$

$$\mathcal{I}_2 = \xi \,^\wedge d\xi = d\psi \,^\wedge dx \,^\wedge dy, \tag{7.7}$$

$$\mathcal{I}_3 = d\xi \,^\wedge d\xi = 0, \tag{7.8}$$

along with the fact that the minimum codimension of the integral submanifolds is two; thus, the maximum dimension will be two, as well.

Since the tangent spaces to the configuration manifold are four-dimensional, and we have one linear constraint in (7.5), the constraint bundle $C(M)$ has three-dimensional fibers; i.e., the motion has three degrees of freedom  One also notes that the fiber coordinate $\omega_N$ is not affected by the constraints, and, in fact, it really represents an independent input to the motion, along with $\omega_T$, since $\omega_N$ is essentially a steering command.  Thus, one can still define integral curves, in the sense of differentiable curves in the $xy$-plane whose velocities are consistent with the constraints, but since the tire can still go from any point of the plane to another one without violating the constraints and can take on arbitrary vertical angles, there are less degrees of freedom to the motion than there are dimensions to the configuration manifold.

## 8. Non-Hamiltonian vector fields.

By now, the way that symplectic structures allow one to define the basic concepts of Hamiltonian mechanics in a manner that is natural to symplectic geometry is becoming widely known to an increasing circle of mathematicians and physicists ([1]).  Recall that a symplectic manifold $(M, \Omega)$ is an even-

---

([1])  It is debatable whether, as Arnol'd once suggested [3], symplectic mechanics has truly obviated the calculus of variations that preceded the transformation from Lagrangian to Hamiltonian methods.  The concepts of action, virtual work, variation (i.e., virtual displacement), and the like are sufficiently natural to physics and engineering that one can hardly imagine why they would be replaced with more esoteric dimensional differentiable manifold $M$ (dim $M = 2m$) that is endowed with a closed non-degenerate 2-form $\Omega$. The most common example of a symplectic manifold in classical mechanics is the cotangent bundle $T^*M$ to any $n$-dimensional differentiable manifold $M$. The 2-form $\Omega$ is then defined by the exterior derivative $d\theta$ of the canonical 1-form $\theta$ on $T^*M$ that we discussed above.

Since non-degeneracy is defined as (or equivalent to) the notion that the linear map $\iota_x$: $T_xM \to T_x^*M$, $\mathbf{X} \mapsto i_{\mathbf{X}}\Omega$ is an isomorphism at every $x \in M$, one can associate every vector field $\mathbf{X}(x)$ with a 1-form $\iota(\mathbf{X})$. Conversely, every 1-form $\xi$ on $M$ can be associated with a unique vector field $\iota^{-1}(\xi)$. In particular, if the 1-form is exact $\xi = dH$ then associated vector field takes the form of the *symplectic gradient* of the function $H$:

$$\nabla_\Omega H = \iota^{-1}(dH). \tag{8.1}$$

One generally calls such vector fields *(globally) Hamiltonian vector fields*. If the 1-form $\xi$ is closed, but not exact, then one refers to the corresponding vector field $\iota^{-1}(\xi)$ as *locally Hamiltonian*. Clearly, the two concepts are identical when $M$ is simply connected, although we shall use the global kind as the customary case.

Since every vector field on a differentiable manifold is locally equivalent to a system of ordinary differential equations, this gives one a means to associate a system of dynamical equations with every differentiable function on $M$, which can be regarded as a Hamiltonian, but can also be regarded as any other physical "observable." This follows from the fact that the Lie bracket of Hamiltonian vector fields closes into a Lie sub-algebra of the Lie algebra $\mathfrak{X}(M)$ of all vector fields on $M$. In fact, the Lie sub-algebra thus defined is "almost isomorphic" to the Lie algebra defined over the smooth functions on $M$ by way of the Poisson bracket:

$$\{f, g\} = -\Omega(\nabla_\Omega f, \nabla_\Omega g) ; \tag{8.2}$$

in particular:

$$[\nabla_\Omega f, \nabla_\Omega g] = \nabla_\Omega\{f, g\}. \tag{8.3}$$

One sees that the linear map $\nabla_\Omega$: $C^\infty(M) \to \mathfrak{X}(M)$, $f \mapsto \nabla_\Omega f$ is a Lie algebra homomorphism that fails to be an isomorphism onto its image due to the fact that constant functions define a one-dimensional Abelian subalgebra of the Poisson algebra that also represents the kernel of the map. One can also regard the Poisson algebra as the "central extension" of the Lie algebra of Hamiltonian vector fields by the constant functions, although we shall have no further use for that fact in the present context.

One should also observe that the local flows of the Hamiltonian vector fields have special properties relating to both the function $H$ that generates them and the symplectic 2-form $\Omega$.

In particular, the function $H$ is constant along the flow of $\nabla_\Omega H$, since the Lie derivative of $H$ with respect to the vector field $\nabla_\Omega H$ is:

---

mathematical constructions that mostly seem intrinsic to point mechanics, while their extension to continuum mechanics and field theories becomes increasingly contrived and non-intuitive.

$$(\nabla_{\Omega} H)\, H = i_{\nabla_{\Omega} H} dH = \Omega(\nabla_{\Omega} H, \nabla_{\Omega} H) = -\{H, H\} = 0. \qquad (8.4)$$

Thus, when the function $H$ is the total energy of the mechanical system this implies that total energy must be conserved during the motion.

One also finds that the 2-form $\Omega$ is constant along the flows, since the Lie derivative of $\Omega$ with respect to $\nabla_{\Omega} H$ is:

$$L_{\nabla_{\Omega} H} \Omega = i_{\nabla_{\Omega} H} d\Omega + d i_{\nabla_{\Omega} H} \Omega = d(dH) = 0. \qquad (8.5)$$

Thus, the local flows of Hamiltonian vector fields are composed of local *canonical transformations* – or *symplectomorphisms*, as they are also called. Such flows also preserve Poisson brackets.

In order to show how all of this relates to the usual Hamiltonian mechanics that most physicists still recognize, one first uses *Darboux's theorem* to find a canonical coordinate chart $(U, x^i, p_i)$ about each point of $M$ in which the symplectic 2-form $\Omega$ takes the form:

$$\Omega = dp_i \wedge dx^i. \qquad (8.6)$$

The matrix of the 2-form $\Omega$ with respect to the natural frame field of a canonical coordinate system then takes the form:

$$\Omega_{IJ} = \begin{bmatrix} 0 & -I \\ I & 0 \end{bmatrix}. \qquad (8.7)$$

Thus, if $H$ is a smooth function on $M$ then the 1-form it defines takes the form:

$$dH = \frac{\partial H}{\partial x^i} dx^i + \frac{\partial H}{\partial p_i} dp_i, \qquad (8.8)$$

and the Hamiltonian vector field takes the local form:

$$\nabla_{\Omega} H = \frac{\partial H}{\partial p_i} \frac{\partial}{\partial x^i} - \frac{\partial H}{\partial x^i} \frac{\partial}{\partial p_i}. \qquad (8.9)$$

The system of $2n$ ordinary differential equations that is then defined by making this vector field take the form of the velocity vector field for any integral curve is then:

$$\frac{dx^i}{dt} = \frac{\partial H}{\partial p_i}, \qquad\qquad \frac{dp_i}{dt} = -\frac{\partial H}{\partial x^i}. \qquad (8.10)$$

Now, the vector field $\nabla_{\Omega} H$ is still a directional derivative operator that acts on the smooth functions on $M$, and one finds that the local form of the Poisson bracket comes about from applying the operator $\nabla_{\Omega} f$ to the function $g$:

$$(\nabla_\Omega f)\, g = \frac{\partial f}{\partial p_i}\frac{\partial g}{\partial x^i} - \frac{\partial f}{\partial x^i}\frac{\partial g}{\partial p_i} = -\{f, g\}. \tag{8.11}$$

Since Hamiltonian functions are automatically associated with the conservation of energy, it might seem that the methods of symplectic geometry become less useful when one goes on to non-conservative mechanical systems. However, one finds that the existence of a symplectic structure $\Omega$ on a manifold $M$, which usually takes the form of a "phase space" in physics, can still be just as useful when one is considering non-conservative dynamical systems.

The key is to regard the vector fields on $M$ as the physically fundamental objects, not the smooth functions. This is essentially equivalent to regarding force as being more fundamental than energy, or 1-forms as more fundamental than 0-forms. Now, let us start with a more general vector field on $M$:

$$\mathbf{X} = X^i \frac{\partial}{\partial x^i} + X_i \frac{\partial}{\partial p_i}, \tag{8.12}$$

and convert it into a 1-form by means of $\Omega$:

$$\xi = i_\mathbf{X}\Omega = -X_i\, dx^i + X^i\, dp_i. \tag{8.13}$$

As we saw in Part I, any 1-form on $M$ can be put into a canonical form:

$$\xi = dH + \sum_{a=1}^{p} \mu_a dv^a, \tag{8.14}$$

in which, as we have seen, $p$ is determined by looking at the sequence of integrability forms $\xi$, $d\xi$, $\xi \wedge d\xi$, …, and it also relates to the maximum dimension of an integral submanifold of $\xi = 0$, namely, $2n - p - 1$.

When one has a 1-form $\xi$ on $M$ that has been expressed in its canonical form, one can conversely, associate it with a vector field using $\Omega$:

$$\mathbf{X}_\xi = \iota^{-1}(\xi) = \nabla_\Omega H + \sum_{a=1}^{p} \mu_a \nabla_\Omega v^a. \tag{8.15}$$

Thus, the symplectic gradient operator behaves somewhat like the exterior derivative, except that one is considering vector fields, instead of 1-forms, so one can think of (8.15) as being the canonical form for an arbitrary vector field $\mathbf{X}$ on $M$ when $\xi = i_\mathbf{X}\Omega$.

One immediately sees from (8.15) that $H$ is not constant along the flow of $\mathbf{X}_\xi$, since:

$$\mathbf{X}_\xi H = \sum_{a=1}^{p} \mu_a \{v^a, H\}. \tag{8.16}$$

Similarly, $\Omega$ is not constant along that flow either, since:

$$L_{X_\xi}\Omega = d(i_{X_\xi}\Omega) = d\left(dH + \sum_a \mu_a dv^a\right) = \sum_a d\mu_a \wedge dv^a \ . \tag{8.17}$$

Thus, the flow of $X_\xi$ does not consist of canonical transformations.

The symplectic gradients $\nabla_\Omega v^\mu$ have local expressions that are analogous to (8.9), so when one considers the local system of ordinary differential equations that is defined by non-Hamiltonian vector field one obtains equations of the form:

$$\frac{dx^i}{dt} = \frac{\partial H}{\partial p_i} + \mu_a \frac{\partial v^a}{\partial p_i}, \tag{8.18}$$

$$\frac{dp_i}{dt} = -\frac{\partial H}{\partial x^i} - \mu_a \frac{\partial v^a}{\partial x^i}. \tag{8.19}$$

The first set of equations defines a generalized velocity vector field on $U$, while the second set defines a generalized force 1-form.

When one takes two arbitrary vector fields $X_\xi$ and $X_\eta$, where:

$$\xi = df + \sum_{a=1}^{p} \mu_a dv^a \ , \qquad \eta = dg + \sum_{b=1}^{p'} \bar{\mu}_b d\bar{v}^b \ , \tag{8.20}$$

and computes the Lie bracket, the result is:

$$[X_\xi, X_\eta] = \nabla_\Omega\{f, g\} +$$
$$+ \sum_{b=1}^{p'} [\nabla_\Omega f, \bar{\mu}_b \nabla_\Omega \bar{v}^b] - \sum_{a=1}^{p} [\nabla_\Omega g, \mu_a \nabla_\Omega v^a] + \sum_{a=1}^{p} \sum_{b=1}^{p'} [\mu_a \nabla_\Omega v^a, \bar{\mu}_b \nabla_\Omega \bar{v}^b] \ . \tag{8.21}$$

It is not difficult to expand the last three Lie brackets into linear combinations of Poisson brackets, although the final expression is quite obese. For instance:

$$\sum_{b=1}^{p'} [\nabla_\Omega f, \bar{\mu}_b \nabla_\Omega \bar{v}^b] = \sum_{b=1}^{p'} \{f, \bar{\mu}_b\} \nabla_\Omega \bar{v}^b + \bar{\mu}_b \nabla_\Omega \{f, \bar{v}^b\} \ . \tag{8.22}$$

The qualitative character of (8.21) is that the Lie algebra that is defined by the Poisson brackets – or the Hamiltonian vector fields – is being deformed by the contributions that come from the non-conservative nature of the flows. Thus, one sees another possible path to quantum mechanics if one can justify that the difference between classical and quantum mechanics relates to the integrability of observables; i.e., their conservation or non-conservation.

**9. Conservation of energy.** The law of balance or conservation of energy has a slightly different character from those of linear or angular momentum, since one generally derives the former from the latter under restrictive assumptions about the nature of the forces and momenta that will allow one to replace them with exact forms. One finds that when one does have conservation of total energy, that fact defines a foliation of one's manifold of kinematical states by the level surfaces of the total energy function $E$, which then become maximal integral submanifolds of the Pfaffian equation $dE = 0$. Thus, one wonders what would happen to that picture if one weakened the restrictions to produce a more general Pfaffian equation.

Let us recall the path from the balance of linear momentum to the balance of energy in the case of point mechanics. One starts with the basic system of ordinary differential equations:

$$F = \frac{dp}{dt}, \qquad (9.1)$$

in which the 1-form $F(x(t))$ represents the resultant force that acts on a point mass whose trajectory is described by a twice-continuously-differentiable curve $x(t)$ in a configuration manifold $M$, and whose linear momentum is described by the 1-form $p(t)$.

One then forms the corresponding equation for the balance of power along the curve by evaluating both sides of (9.1) on the velocity vector field $\mathbf{v}(t)$:

$$F(\mathbf{v}) = \frac{dp}{dt}(\mathbf{v}). \qquad (9.2)$$

One then assumes that linear momentum is coupled to velocity by a linear constitutive law:

$$p = m(\mathbf{v}) \qquad (p_i = m_{ij}v^j), \qquad (9.3)$$

in which one assumes that the mass matrix $m_{ij}$ is symmetric, invertible, and time-invariant. This is the usual case when one uses $m_{ij} = m \, \delta_{ij}$, where $m$ is then the mass of the particle and represents the components of a Euclidian metric on $T(M)$, as described in an orthonormal frame. However, an anti-symmetric contribution to $m_{ij}$ is not out of the question, as it produces a "transverse momentum" – i.e., a component that is not collinear with covelocity, and such a concept does appear in the dynamics of the Dirac electron, as well as its approximation in the form of the Weyssenhoff fluid.

With these restrictions, one can say that:

$$\frac{dp}{dt}(\mathbf{v}) = \frac{dT}{dt}, \qquad (9.4)$$

in which we have set:

$$T = \tfrac{1}{2} p(\mathbf{v}) = \tfrac{1}{2} \, m_{ij} \, v^i \, v^j, \qquad (9.5)$$

which then represents the non-relativistic kinetic energy of the particle.

One then forms the 1-forms that are defined by multiplying both sides of (9.2), with the substitution (9.4), by $dt$ and then integrating along a curve segment $\gamma$ that goes from $t_0$ to $t_1$:

$$\int_{t_0}^{t_1} F(\mathbf{v})\, dt \;=\; \int_{T_0}^{T_1} dT = \Delta T[t_0, t_1] \qquad\qquad (9.6)$$

Since $v^i = dx^i / dt$, the left-hand side becomes:

$$\Delta U[\gamma] = \int_\gamma F \;, \qquad\qquad (9.7)$$

but this will generally depend upon the path, depending upon the nature of $F$. It is only if $F$ is an exact 1-form – say – $dU$ – that $\Delta U[\gamma]$ can be replaced with $\Delta U[t_0, t_1]$, and one can then assert that:

$$\Delta E = 0, \qquad\qquad (9.8)$$

in which we have defined the total energy of the particle by:

$$E = T + U. \qquad\qquad (9.9)$$

We are then dealing with the conservation of total energy.

Otherwise, if one has a more general normal form for $F$, such as $- dU + \mu_a\, dv^a$, then one can only assert the balance of total energy in the form:

$$\Delta E = \int_\gamma \mu_a dv^a \;, \qquad\qquad (9.10)$$

which will then be path-dependent.

One can reset this same situation on the manifold of kinematical states for point particle motion, namely, $J^1(\mathbb{R}; M)$, by defining a generalized force 1-form on it:

$$\mathcal{F} = F_i\, dx^i + p_i\, dv^i, \qquad\qquad (9.11)$$

in which the functional relationships:

$$F_i = F_i(t, x, v), \qquad\qquad p_i = p_i(t, x, v) \qquad\qquad (9.12)$$

represent the empirical constitutive laws that specify the nature of the forces and momentum.

In the case of a conservative force and an exact kinematical force term $p_i\, dx^i = dT$, one can say that:

$$\mathcal{F} = d\mathcal{L}, \qquad\qquad (\mathcal{L} = T - U), \qquad\qquad (9.13)$$

in which $\mathcal{L} = \mathcal{L}(x, v)$ then plays the role of a Lagrangian function for the action functional that is associated with this motion.

In this situation, the Pfaffian equation $\mathcal{F} = 0$ is completely integrable, and the maximal integral submanifolds are simply the level surfaces of $\mathcal{L}$. However, when one has non-conservative forces $F_i \, dx^i$ will not be exact, and the presence of transverse momenta can render $p_i \, dv^i$ inexact, as well. In order to see this, suppose that one has (9.3) with no restriction on the symmetry of the mass matrix. One then has:

$$d(p_j \, dv^j) = d(m_{ij} \, v^i \, dv^j) = \tfrac{1}{2}(m_{ij} - m_{ji}) \, dv^i \wedge dv^j. \tag{9.14}$$

If $m_{ij}$ is not symmetric then this 2-form will not vanish, and one will generally have:

$$p_i \, dv^i = dT + \rho_b \, d\sigma^b \ . \tag{9.15}$$

Thus, one must generally consider a Pfaff equation $\mathcal{F} = 0$, in which the normal form for $\mathcal{F}$ does not consist of merely an exact form, and whose maximal integral submanifolds are of higher codimension than one.

Dually, if one considers the 1-form:

$$\mathcal{F} = -F_i \, dx^i + v_i \, dp^i \tag{9.16}$$

on the manifold $J^1(M; \mathbb{R})$ then if it is exact (recall that we are assuming that $m_{ij}$ is invertible), the Pfaff equation $\mathcal{F} = 0$ takes the form $dH = 0$, where we have defined the Hamiltonian function $H = T + U$. One can then say that the maximal integral surfaces represent the conservation of energy. That is, a section of the *target* projection $J^1(M; \mathbb{R})$ $\to \mathbb{R}$ must lie in such a hypersurface in order for there to be conservation of energy.

In the more general case, of an inexact $\mathcal{F}$, one finds that conservation of energy becomes more restrictive, and one generally contends with the balance of energy, instead.

## 10. Equilibrium thermodynamics.

**10. Equilibrium thermodynamics.** One of the earliest applications of Pfaff's theory to thermodynamics was by Clausius in 1867 [**4**]. Somewhat later, in 1905, Carathéodory [**5**] made the theory of the Pfaff problem, including the later work of Frobenius, a key part of the definition of entropy in his own axiomatization of thermodynamics. Basically, one finds that one is trying to refine the definition of the force 1-form that gives the work done along a curve from something that is exact to something that has extra inexact terms added to it, as in the canonical form for a Pfaffian form. However, in order for the manifold in which the curve exists to still be finite-dimensional, one must restrict oneself to equilibrium thermodynamics.

When the atomic or molecular constituents of a physical system are in a state of thermal equilibrium, there is no flow of energy from one point to another and the various thermodynamic variables, such as temperature, pressure, density, entropy and the like, cease to be represented by functions on the space occupied by the system that will depend upon the position and time point examined and can be represented as simply constants. Thus, spaces of thermodynamic variables, such as $(P, V, T)$, for example, really represent the "field space" for functions on the spatial domain that is occupied by the system. As a consequence, all of the points in such spaces represent equilibrium states, since every constant value defines a unique constant function when the domain is given.

One also commonly introduces a notion of "quasi-static" equilibrium into equilibrium thermodynamics in order to make the variation of thermodynamic variables in time, or at least, during some sort of "process," take the form of a curve in the space of thermodynamic variables, instead of a curve in the space of thermodynamic functions on the spatial region.

Ultimately, a space of thermodynamic variables becomes a subset of a vector space that is bounded by constraints that usually take the form of non-negativity of the variables. More specifically, the vector space often involves a pairing of a vector space and its dual, if only as a subspace. For instance, consider the space $\Gamma = (U, P, S, V, T)$, whose variables are the total internal energy $U$ in the system, which one should think of as the total of all *modeled* forms of energy, the pressure $P$, entropy $S$, volume $V$, and temperature $T$. One can then think of the thermodynamic variables as coordinate functions on the five-dimensional differentiable manifold (with boundary) $\Gamma$. As we shall see, it can also be regarded as a contact manifold in various ways.

Although the first law of thermodynamics is often referred to as the conservation of energy, actually, the fundamental problem of thermodynamics is to resolve the *non-conservation* of energy, in the form of the work $W[c_1] = \int_{c_1} F$ that is done *by* a physical system along a singular 1-chain $c_1$ (i.e., a curve segment), into its *conservation*, by expanding the state space of the system to include more thermodynamic variables, even though sometimes it seems as if the arguments start to sound tautological.

For instance, one can *define* the change in heat contained in the system by way of:

$$\Delta Q = W[c_1] + \Delta U. \tag{10.1}$$

in which $c_1$ merely defines the curve in the state space that corresponds to the process in question.

Since the kernel of the integral that defines $W[c_1]$ – viz., $F$, or $\,\rlap{/}d W$ , as it is sometimes denoted in thermodynamics – is generally inexact, while the kernel of $\Delta U$ – viz., $dU$ – is exact, one must have the kernel of $\Delta Q$ must also be inexact, so we denote it by $\,\rlap{/}d Q$ . The differential form of (10.1) can then be put into the form:

$$\rlap{/}d W = -\, dU + \rlap{/}d Q, \tag{10.2}$$

which then suggests that what one is really doing at the fundamental level is trying to resolve an inexact 1-form that accounts for the work done on a system into its canonical form as a Pfaffian form, when one regards the exact part as describing the changes in modeled forms of energy – e.g., kinetic and potential – while the other terms successively refine the specific nature of the unmodeled forms. Although it is traditional to refer to $dW$ as a "differential increment of work," we shall simply refer to it as the force 1-form. One sees that the non-integrability of work in thermodynamics has much do with the incompleteness in the model for the physical system.

For instance, if two moving automobiles crash head-on and both come to a stop then if the total energy is merely the sum of the kinetic energies, one must say that it is not conserved. However, one could introduce the concept of "binding energy" as a way of saying that total energy is conserved, after all. Of course, it would be more fundamental to make binding energy into a sum of kinetic and potential energies for some more detailed picture of the vehicles. For instance, one could treat them as extended material structures that were deformed and heated by the collision, so the binding energy becomes the sum of the work done deforming both structures, as well as the amount of heating of the materials. One could then resolve the work done deforming into a change in potential energy at the atomic level, while the change in heat was essentially the change in total kinetic energy of the constituent atoms.

Of course, one must notice that the dimension of the state space keeps growing explosively at each step. In effect, the thermodynamic considerations amount to a sort of dimensional "truncation error" in the dynamical model for the states of the system.

The concept of work done around loops is of fundamental significance in equilibrium thermodynamics, such as the *Carnot cycles* that one defines in *PV-space*. Such cycles are composed of four sequential path segments that are alternately isothermal and adiabatic. Thus, a loop in a thermodynamic space represents a cyclic process.

*a. The canonical form for $dW$.* Interestingly, in order to resolve the nebulous term $dQ = dU + dW$ into more detailed contributions, it helps to discuss the second law of thermodynamics before the first law. Basically, one considers the degree of integrability of the inexact 1-form $dQ$. The most elementary form it can take beyond exactness is an exact 1-form times an integrating factor:

$$dQ = T\, dS. \tag{10.3}$$

Actually, if the functions $T$, $S$ are to be interpreted as absolute temperature and entropy then the choice of $T\, dS$, instead of $S\, dT$, is dictated by the nature of the system. The former choice refers to an isothermal system while the latter one, to an isentropic system.

One can also encounter higher-dimensional possibilities that involve higher degrees of integrability for $dQ$. Commonly, for a chemical system one might consider that the molecules in the system might be divided into $s$ different interacting species with total populations $n_a$, $a = 1, \ldots, s$, so the discrepancy on the work functional might be due to both changes in temperature/entropy or changes in numbers of molecules of each type, which might put $dQ$ into the form:

$$dQ = T\, dS + \sum_{a=1}^{s} \mu_a dn^a. \tag{10.4}$$

The $\mu_a$ are referred to as the *chemical potentials*, since they have the units of energy and relate to the interaction energies associated with the changes in numbers of the molecules due to conversions from one species to another – i.e., reactions.

Ultimately, one obtains a force 1-form such as:

$$\not{d}W = - \, dU + T \, dS + \sum_{a=1}^{s} \mu_a dn^a \, , \qquad (10.5)$$

and we can use this 1-form to define a Pfaffian form:

$$\theta = \not{d}W + dU - T \, dS - \sum_{a=1}^{s} \mu_a dn^a \, . \qquad (10.6)$$

Once we have refined our definition of $\not{d}W$, we can show that this is the canonical 1-form on a contact manifold that makes the first law of thermodynamics essentially a non-holonomic linear constraint on that manifold.

The usual form that the work done on a thermodynamic systems takes is by way of applying a pressure $P$ to the volume $V$ to produce an increment $W[\Delta V] = - \, P \, \Delta V$, in which the negative sign is based in the fact that it takes positive work to compress a volume. As it is physically unrealistic to think of pressure staying constant while the volume changes, one can give this law a differential formulation instead:

$$\not{d}W = - \, P \, dV, \qquad (10.7)$$

which makes the Pfaffian form look like:

$$\theta = dU + P \, dV - T \, dS - \sum_{a=1}^{s} \mu_a dn^a \, . \qquad (10.8)$$

We can regard (10.8) as the canonical 1-form for the contact manifold $J^1(\Sigma; \mathbb{R})$, where $\Sigma$ is the subset of $\mathbb{R}^{2+s} = (V, S, n^a)$ that consists of all non-negative coordinates, $U = U(V, S, n^a)$, and the coordinates of $J^1(\Sigma; \mathbb{R})$ can then take the form $(V, S, n^1,\dots, n^s, U, P, - T, \mu_1, \dots, \mu_s)$. For an integrable section $s: \Sigma \rightarrow J^1(\Sigma; \mathbb{R})$ of the source projection, which locally looks like:

$$s(V, S, n_a) = (V, S, n^a, U(V, S, n^a), P(V, S, n^a), - T(V, S, n^a), \mu_a(V, S, n^a)),$$

one will then have:

$$0 = s^* \theta = dU + P \, dV - T \, dS - \sum_{a=1}^{s} \mu_a dn^a \qquad (10.9)$$

i.e., conservation of energy.

The necessary and sufficient condition for the integrability of $s$ can also be given by the vanishing of the Spencer derivative of $s$:

$$Ds = s - j^1 U = -\left(P + \frac{\partial U}{\partial V}\right) dV + \left(T - \frac{\partial U}{\partial S}\right) dS + \left(\mu_a - \frac{\partial U}{\partial n^a}\right) dn^a, \qquad (10.10)$$

which is a 1-form on $\Sigma$ ($^1$). If this vanishes then one can conclude that:

$$P = -\frac{\partial U}{\partial V}, \qquad T = \frac{\partial U}{\partial S}, \qquad \mu_s = \frac{\partial U}{\partial n_a}. \qquad (10.11)$$

The exterior derivative operator commutes with pull-backs:

$$s^* \Omega = s^* d\theta = d(s^* \theta),$$

so if $s^* \theta$ vanishes then so will $s^* \Omega$. Thus, one sees that an integrable section of the source section is a Legendrian submanifold of the contact manifold $J^1(\Sigma, \mathbb{R})$, which is then another way of characterizing the first law of thermodynamics.

One can immediately compute:

$$\Omega = dp \wedge dV - dT \wedge dS - \sum_{a=1}^{s} d\mu_a \wedge dn^a. \qquad (10.12)$$

We now assume that there are no chemical reactions going on, so if $\Sigma$ is the positive quadrant of $\mathbb{R}^2 = (V, S)$ then the Pfaffian form on $J^1(\Sigma; \mathbb{R})$ reduces to:

$$\theta = dU + P\, dV - T\, dS. \qquad (10.13)$$

The algebraic solution to the exterior differential equation $\theta = 0$ then consists of a hyperplane in the tangent space at each point in $J^1(\Sigma; \mathbb{R})$ that defines a constraint on the possible tangent vectors to the curves in $J^1(\Sigma; \mathbb{R})$ if one is to be consistent with conservation of energy. However, the maximal integral submanifolds will be two-dimensional, not four-dimensional. One can see this by examining the successive expressions in the integrability sequence:

$\mathcal{I}_1 \colon \Omega \qquad = dP \wedge dV + dS \wedge dT,$

$\mathcal{I}_2 \colon \theta \wedge \Omega \qquad = dU \wedge dP \wedge dV + dU \wedge dS \wedge dT + P\, dV \wedge dS \wedge dT + S\, dT \wedge dP \wedge dV,$

---

($^1$)  Since the target space in this case is simply $\mathbb{R}$, we are omitting the explicit reference to a vector field on $\mathbb{R}$ in $Ds$.

$\mathcal{I}_3$: $\Omega \wedge \Omega$        $= 2\, dP \wedge dV \wedge dS \wedge dT,$

$\mathcal{I}_4$: $\theta \wedge \Omega \wedge \Omega$  $= 2\, dU \wedge dP \wedge dV \wedge dS \wedge dT.$

Now, the last expression is proportional to the volume element on $\Gamma$, which is then non-vanishing, although the next term $\Omega \wedge \Omega \wedge \Omega$, being a 6-form on a five-dimensional manifold, must vanish automatically. Thus, the minimum codimension of an integral submanifold in $J^1(\Sigma; \mathbb{R})$ is three, which makes the maximum dimension two. Hence, the maximal integral submanifolds are Legendrian submanifolds. In fact, a good choice of integral submanifold is given by $j^1 U: \Sigma \to J^1(\Sigma; \mathbb{R})$, $(V, S) \mapsto j^1 U(V, S)$, with:

$$j^1 U(V, S) = (V, S, U(V, S), \partial U / \partial V, - \partial U / \partial S). \tag{10.14}$$

Before continuing, we point out that it is by using the fact that on an integral submanifold $s: \Sigma \to J^1(\Sigma; \mathbb{R})$, one must have $s^*\Omega = 0$ that one obtains the various Maxwell identities. Namely, if $P = P(V, S)$ and $T = T(V, S)$ then:

$$dP = \frac{\partial P}{\partial V} dV + \frac{\partial P}{\partial S} dS \,, \qquad dT = \frac{\partial T}{\partial V} dV + \frac{\partial T}{\partial S} dS \,, \tag{10.15}$$

and

$$\Omega = \left( \frac{\partial T}{\partial V} + \frac{\partial P}{\partial S} \right) dV \wedge dS \,. \tag{10.16}$$

The vanishing of $s^*\Omega$ must then imply that:

$$\frac{\partial T}{\partial V} = - \frac{\partial P}{\partial S} \,, \tag{10.17}$$

which is the first Maxwell identity. The other three are obtained analogously by using the other three forms for $\theta$, which we shall define shortly.

Thus, one sees that the Maxwell identities are the necessary and sufficient conditions for conservation of energy to be true on any Legendrian submanifold.

There is nothing special about using $(V, S)$ as the basic variables, and in fact, the choice is usually dictated by the nature of the thermodynamic system and its *thermodynamic potential* function. We have chosen the internal energy $U$ as a function on $(V, S)$, but other popular choices are:

*Enthalpy:*
$$H(S, P) = U + PV, \tag{10.18}$$

*Helmholtz free energy:*
$$F(T, V) = U - TS, \tag{10.19}$$

*Gibbs free energy:*

$$G(T, P) = U - TS + PV. \qquad (10.20)$$

The types of systems that these choices are appropriate to generally take the form of systems in which the variables that do not appear as independent variables are regarded as being constant. For instance, internal energy is used as a thermodynamic potential for systems in which the pressure and temperature are regarded as constant, enthalpy, for systems in which temperature and volume are constant, etc. Thus, the corresponding contact form $\theta$ is associated with each of the last three thermodynamic potentials is:

$$dH - T \, dS - V \, dP, \qquad dF + S \, dT + P \, dV, \qquad dG + S \, dT + V \, dP, \qquad (10.21)$$

resp., while the expressions for the conjugate variables as partial derivatives of the thermodynamic potentials, under the assumption that conservation of energy is fulfilled, are:

$$T = \frac{\partial U}{\partial S} = \frac{\partial H}{\partial S}, \qquad V = \frac{\partial H}{\partial P} = \frac{\partial G}{\partial P}, \qquad (10.22)$$

$$P = -\frac{\partial U}{\partial V} = -\frac{\partial F}{\partial V}, \qquad S = -\frac{\partial F}{\partial T} = -\frac{\partial G}{\partial T}. \qquad (10.23)$$

Thus, in any case, we are looking at the manifold $J^1(\Sigma, \mathbb{R})$, where $\Sigma$ is some region in $\mathbb{R}^2$, which often amounts to the non-negative quadrant. The function on $\Sigma$ is then defined by one of the thermodynamic potentials, and as a result, the derivative coordinates are paired with the coordinates of $\Sigma$ as conjugate variables. The canonical 1-form $\theta$ on $J^1(\Sigma, \mathbb{R})$ gives the first law of thermodynamics in the form of the exterior differential system $\theta = 0$. Hence, the acceptable sections $\phi$ of the source projection $J^1(\Sigma, \mathbb{R}) \to \Sigma$ are the ones for which $\phi^* \theta = 0$; i.e., Legendrian submanifolds. In any case, the canonical 2-form $d\theta$ always has the same form, which is $dP \wedge dV + dT \wedge dS$. We summarize the four popular possibilities in Table 3 below.

Table 3. Contact manifolds associated with common thermodynamic potentials.

| Thermodynamic potential | Contact manifold | Canonical 1-form |
|:---:|:---:|:---:|
| $U$ | $(V, S, U, P, -T)$ | $dU + P \, dV - T \, dS$ |
| $F$ | $(V, T, F, P, S)$ | $dF + P \, dV + S \, dT$ |
| $H$ | $(P, S, H, -V, -T)$ | $dH - V \, dP - T \, dS$ |
| $G$ | $(P, T, U, -V, S)$ | $dG - V \, dP + S \, dT$ |

In any case, one will have that $d\theta$ takes the form that was given above, and that the necessary and sufficient conditions for the surface in $J^1(\Sigma; \mathbb{R})$ that is defined by conservation of energy to be a Legendrian submanifold are the Maxwell identities that are associated with the form $\theta$.

*b. Carathéodory's theory.* One of the most debated aspects of Carathéodory's axiomatic formulation of equilibrium thermodynamics [**5**] (cf., also Pippard [**6**] and Buchdahl [**7**]) is his way of making the existence of the functions $T$ and $S$ equivalent to not only the vanishing of the Frobenius 3-form $d\!\!\!/Q \wedge d(d\!\!\!/Q)$ but to what some have called the principle of *local adiabatic inaccessibility.*

First, we shall say that a quasi-static thermodynamic process $\gamma: \mathbb{R} \rightarrow \Gamma$ in the manifold $\Gamma$ of equilibrium states – i.e., a smooth curve – is *adiabatic* iff it is consistent with the constraint that is imposed by $d\!\!\!/Q = 0$, so:

$$d\!\!\!/Q(\dot{\gamma}) = 0 \qquad \text{for all } t. \qquad (10.24)$$

Thus, there is no change of the total heat content at every stage of the process. One can also say that $\gamma$ is an integral curve for the Pfaffian equation $d\!\!\!/Q = 0$ when the process that it describes is adiabatic.

If $x \in \Gamma$ is an equilibrium state then the state $y \in \Gamma$ is *adiabatically accessible* iff there exists an adiabatic process $\gamma(t)$ such that $x = \gamma(0)$ and $y = \gamma(1)$; otherwise, $y$ is adiabatically inaccessible from $x$. The way that Carathéodory phrased his theorem assumed that $\Gamma$ was an (ill-defined) region of $\mathbb{R}^n$ – or possibly all of it – and he asserted if every neighborhood $U$ of every $x$ contained at least one state that was adiabatically inaccessible from $x$ then there existed functions $T$, $S$ (of unspecified domain and regularity) such that $d\!\!\!/Q = T\,dS$. Somewhat later, Bernstein [**8**] refined the theorem by emphasizing that it really had a local character to it, so what he proved was that if $d\!\!\!/Q$ were a smooth 1-form on some well-defined region $G \subset \Gamma$ and was positive-definite on $G$ – in the sense that $\sum_{i=1}^{n} X_i X_i > 0$ – and if the curve $\gamma$ were piecewise smooth in $U$ then one could conclude the existence of such smooth functions $T$, $S$ only for some neighborhood of $x$ that is contained in $U$. Boyling [**9**] then refined the theorem further by letting $\Gamma$ be a smooth differentiable manifold without boundary while $d\!\!\!/Q$ was a globally non-vanishing smooth 1-form on $\Gamma$ and then proving:

**Theorem:**

*The following are equivalent:*

1. *For every $x \in G$ there exists an open neighborhood $U$ of $x$ such that there are smooth functions $T$, $S$ on $U$ that make the restriction of $d\!\!\!/Q$ to $U$ take the form $T\,dS$.*

2.   $dQ \wedge d(dQ) = 0$.

3.   *For every $x \in G$ there exists an open neighborhood $U$ of $x$ such that every neighborhood $V$ of $x$ that is contained $U$ also contains an adiabatically inaccessible point $y$.*

The equivalence of the first two statements is basically a consequence of the canonical form for the 1-form $dW$ in three dimensions. The extension of this local existence to a global existence is much more involved and gets into the predictable topological considerations.

For more recent discussions of Carathéodory's axioms, one can confer Rajeev [**10**], as well as Cheng and Tseng [**11**].

**11. Vortical flow.** The next step up from a 1-form being closed, but not exact, is for it to be not closed, either. Of course, as we just saw in the preceding section, this possibility can be refined further by establishing the degree of integrability of the 1-form in question, and thus, the canonical form.

In the theory of vortical flow (see, e.g., [**12-15**]), one key 1-form to examine on the space or spacetime manifold $M$ – at least, the part in which the fluid is defined – is the *flow covelocity* 1-form $v = v_i \, dx^i$ that is dual to the *flow velocity* vector field $\mathbf{v} = v^i \, \partial_i$ by way of the metric tensor $g = g_{ij} \, dx^i \, dx^j$ that is defined on $M$; thus, $v_i = g_{ij} \, v^j$.

Another 1-form that is closely related to the covelocity is the *momentum 1-form* $p = p_i \, dx^i$, whose relationship to $v$ or $\mathbf{v}$ might be as simple as $p = \rho \, v$, where $\rho$ is the mass density, or a more general constitutive law $p = \rho(\mathbf{v})$, for which $\rho : \Lambda_1 M \to \Lambda^1 M$ is an invertible bundle map that takes vector fields to 1-forms.

If we consider the 1-form $v$ then the obvious problem to pose is the Pfaff problem. The exterior equation $v = 0$ basically associates a hyperplane $\text{Ann}[v]_x$ in $T_x M$ to each point $x \in M$ where $v$ is defined. The integrability of $v$ then amounts to the maximum dimension of integral submanifolds that one can find in $M$ whose tangent vectors lie in the vector sub-bundle $\text{Ann}[v] \to M$. In the completely integrable case, the submanifold has codimension one, as does $\text{Ann}[v]$.

Now, a tangent vector $\mathbf{X}$ to an integral submanifold must satisfy:

$$0 = v(\mathbf{X}) = g(\mathbf{v}, \mathbf{X}), \tag{11.1}$$

which is to say, it must be orthogonal to the velocity vector field. Hence, in the completely integrable case the trajectories (viz., path-lines or streamlines) of the velocity vector field must be intersected orthogonally by hypersurfaces; one then refers to such a flow as *hypersurface orthogonal*.

As we saw above, the two possibilities for complete integrability are that $v$ is exact, so $v = d\psi$, for some *velocity potential function* $\psi$, or that the Frobenius 3-form $v \wedge dv$ must vanish, even though $dv$ does not have to. The 2-form $\Omega_k = dv$ is sufficiently fundamental as to deserve a name, and one calls it the *kinematical vorticity* 2-form, as distinguished from the *dynamical vorticity* $\Omega_d = dp$, which does not have to be proportional to $\Omega_d$, depending upon the nature of the constitutive law. Thus, one can also represent the Frobenius 3-form as:

$$v \wedge \Omega_k = \tfrac{1}{3}(v_i \Omega_{jk} + v_j \Omega_{ki} + v_k \Omega_{ij}) \, dx^i \wedge dx^j \wedge dx^k. \tag{11.2}$$

*a. Non-relativistic fluid motion.* Now, in the case of a three-dimensional (i.e., spatial) $M$, which is appropriate to non-relativistic hydrodynamics, a 2-form, such as $\Omega_k$ must be decomposable, so there is some (non-unique) pair of 1-forms $\alpha$ and $\beta$ on $M$ such that $\Omega_k = \alpha \wedge \beta$. Since $\alpha$ and $\beta$ must be linearly independent in order for $\Omega_k$ to be non-vanishing, they must span a 2-plane $[\Omega_k]$ in each cotangent space $T_x^*(M)$, and the vanishing of $v \wedge \Omega_k$ is equivalent to the condition that the covector $v|_x$ must be incident on this plane at that point. In other words, when the flow is vortical – so $dv \neq 0$ – complete integrability of the Pfaff equation $v = 0$ is equivalent to saying that the covelocity is incident on the vorticity planes.

Another way of representing the Frobenius condition is to assume that $M$ is orientable and endowed with a unit-volume element:

$$V = dx^1 \wedge dx^2 \wedge dx^3 = \tfrac{1}{3!} \varepsilon_{ijk} \, dx^i \wedge dx^j \wedge dx^k, \tag{11.3}$$

which is unavoidable if one is to consider the issue of compressibility, then the Poincaré isomorphism #: $\Lambda_1 M \to \Lambda^2 M$ associates every tangent vector $\mathbf{X} = X^i \, \partial_i$ with a 2-form:

$$\#\mathbf{X} = i_{\mathbf{X}} V = \tfrac{1}{2} \varepsilon_{ijk} X^k \, dx^i \wedge dx^j, \tag{11.4}$$

and conversely, any 2-form, such as $\Omega_k$, with a vector field:

$$\boldsymbol{\omega}_k = \#^{-1} \Omega_k = \tfrac{1}{2} \varepsilon^{ijk} \Omega_{jk} \, \partial_i. \tag{11.5}$$

that one thus calls the *kinematic vorticity vector field.*

One finds that the Frobenius 3-form can now be expressed in the form:

$$v \wedge \Omega_k = v \wedge \#\boldsymbol{\omega}_k = v(\boldsymbol{\omega}_k) \, V = g(\mathbf{v}, \boldsymbol{\omega}_k) \, V. \tag{11.6}$$

Thus, complete integrability of $v$ can then be expressed in the equivalent forms of saying that the kinematical vorticity vector field must be incident on the planes defined by the covelocity or orthogonal to the velocity vector field.

When $v = 0$ is completely integrable and $v$ is not closed, one then sees that $v$ must take the form:

$$v = \mu \, d\lambda. \tag{11.7}$$

The maximal integral submanifolds be the local surfaces of $\lambda$.

The third possibility when $M$ is three-dimensional is that $v \wedge \Omega_k$ does not vanish, since one must have the vanishing of $dv \wedge dv = \Omega_k \wedge \Omega_k$ on dimensional grounds. In such a case, one must have:

$$v = d\psi + \mu \, d\lambda . \tag{11.8}$$

In the context of vortical flow, one still refers to the function $\psi$ as the velocity potential, while the functions $\mu$, $\lambda$ are the *Clebsch variables*.

Since this third possibility represents incomplete integrability of the Pfaff equation $v = 0$, one finds that no orthogonal surfaces to the streamlines will exist. Thus, the maximal integral submanifolds will be curves. However, the non-vanishing of $v \wedge dv$ is equivalent to the non-vanishing of $v(\boldsymbol{\omega}_k)$ or $g(\mathbf{v}, \boldsymbol{\omega}_k)$, which means the trajectories of $\boldsymbol{\omega}_k$ (i.e., the *vortex lines*) will not be incident on the planes defined by $v$ or orthogonal to the streamlines.

### b. Relativistic fluid flow.

When the manifold $M$ is four-dimensional, one can further distinguish between the four-dimensional Minkowski space ($\mathbb{R}^4$, $\eta_{\mu\nu}$) of special relativity and the more general Lorenzian manifolds ($M$, $g$) of general relativity. The main geometrical difference between them is that in Minkowski space, one can use the natural parallel translation that makes any vector space into an affine space, while in a more general Lorentzian manifold, the parallel translation of tangent and cotangent objects must be chosen from a broad palette of possibilities and is usually only possible locally about any given point.

The next consideration, beyond the dimension of $M$, that one must address for the modeling of relativistic hydrodynamics [16-18] is the fact that the indefinite character of Lorentzian metrics allows one to distinguish between three types of tangent vectors. If we choose the signature convention that $\eta_{\mu\nu} = \mathrm{diag}[+1, -1, -1, -1]$ then a *timelike* tangent vector $\mathbf{v}$ satisfies $g(\mathbf{v}, \mathbf{v}) > 0$, a *light-like* one satisfies $g(\mathbf{v}, \mathbf{v}) = 0$, and a *spacelike* one satisfies $g(\mathbf{v}, \mathbf{v}) < 0$. Massive matter must move on timelike trajectories, massless matter must move on lightlike ones, and generally spacelike curves are regarded as unphysical, since they would be associated with superluminal velocities and imaginary proper time elements.

As far as the integrability of the Pfaff equation $v = 0$ is concerned, the main difference is in the extra dimension on the manifold $M$. As a result, there will be non-vanishing 4-forms, and in particular, $dv \wedge dv$ does not have to vanish, although $v \wedge dv \wedge dv$ still does.

One consequence of this is the fact that the 2-form $\Omega_k = dv$ does not have to be decomposable, so there are two types of non-zero 2-forms on a four-dimensional manifold:

$$\Omega_k = \alpha \wedge \beta \qquad \text{and} \qquad \Omega_k = \alpha \wedge \beta + \rho \wedge \sigma,$$

which correspond to the case in which $\Omega_k \wedge \Omega_k$ does and does not vanish, respectively.

The extra canonical form for $v$ that comes into play because of the extra dimension is:

$$v = \mu_1 \, d\lambda_1 + \mu_2 \, d\lambda_2 , \tag{11.9}$$

which describes the cases in which $\Omega_k \wedge \Omega_k$ is non-vanishing.

One must also note that the Poincaré isomorphism now takes the form #: $\Lambda_k M \to \Lambda^{4-k} M$, so the 2-form $\Omega_k$ is not associated with a vector field, any more, but a bivector field. In fact, it is the Frobenius 3-form $v \wedge \Omega_k$ that becomes the Poincaré dual of a vector field:

$$v \wedge \Omega_k = \#\boldsymbol{\omega}_k \tag{11.10}$$

that plays the role of a relativistic (kinematic) vorticity vector field.

One can then verify that the velocity vector field is always orthogonal to the vorticity vector field since:

$$g(\mathbf{v}, \boldsymbol{\omega}_k) \, V = v(\boldsymbol{\omega}_k) \, V = v \wedge \# \, \boldsymbol{\omega}_k = v \wedge v \wedge \Omega_k = 0.$$

It is also clear that there will be three possible dimensions for integral submanifolds in four dimensions, namely, 3, 2, and 1. The first possibility is complete integrability, for which $v$ takes the form of either $d\psi$ or $\mu \, d\lambda$, the second possibility corresponds to $v = d\psi + \mu \, d\lambda$, and the last one, to $v$ that have the form (11.9).

**12. $U(1)$ gauge field theories.** Although the concept of a connection on a fiber bundle, such as a principal fiber bundle, generally gets into the realm of Pfaffian systems, since one is usually dealing with more than one 1-form, nevertheless, when the structure group of the bundle is a one-dimensional Lie group, such as $U(1)$, a connection is associated with a single 1-form. In such cases, the theory of the Pfaff equation becomes directly applicable.

The $U(1)$ gauge field theory that is the best established by experimental confirmation is the gauge formulation of electromagnetism. That story begins when one takes one of the two Maxwell equations, namely, $dF = 0$, to mean that the electromagnetic field strength 2-form $F$ is a closed 2-form on the four-dimensional spacetime manifold $M$. One can also interpret this equation as saying that the forces that go into the definition of $F$ are conservative forces. One thus sees that if one takes the more physically tenable position that these forces are more fundamental than the potential functions and 1-forms (i.e., vector potentials) that might represent these forces (on the assumption that they are, indeed, conservative) then there are some immediate issues that must be addressed:

1. Does every closed 2-form, such as $F$, admit a *global* electromagnetic potential 1-form $A$:

$$F = dA; \tag{12.1}$$

i.e., is every closed 2-form on $M$ an exact 2-form?

Of course, by de Rham's theorem, this would be equivalent to the vanishing of real singular cohomology vector space $H^2(M; \mathbb{R})$ of $M$ in two dimensions, so this is an unavoidably topological question regarding $M$. If there are real 2-cycles in $M$ that do not bound real 3-chains then potential 1-forms can exist only locally. This possibility eventually leads to possible existence of "magnetic monopoles," although the experimental searches for such particles have been unpromising, to date.

2. Whether $A$ exists locally or globally, in either case, it will not be unique, since one can add any closed 1-form $\alpha$ to $A$ and still produce the same $F$ under exterior differentiation:

$$d(A + \alpha) = dA = F. \tag{12.2}$$

As long as one is dealing locally, one can also treat all closed 1-forms as being exact, so one can set $\alpha = d\lambda$ for some smooth function $\lambda$, which is then defined only up to a constant, its own right ([1]), and which one refers to as a choice of gauge for $F$, although perhaps it is more precise to say that the pair $\{A, \lambda\}$ is a gauge. The 2-form $F$ is then insensitive to the replacement:

$$A \rightarrow A + d\lambda, \tag{12.3}$$

which one then refers to as a *gauge transformation of the second kind.*

In order to explain the gauge transformations of the first kind, one must go further into the interpretation of $\lambda$. The leap of faith comes from saying that if $A$ is really the local representative of a connection 1-form on a $G$-principal bundle $P \rightarrow M$ over $M$ for which the structure group $G$ is one-dimensional then the formula (12.3) would take the form of the transformation of the values of the local connection 1-form $A$ on some $U \subset M$ to the corresponding representative $A'$ that is defined on some overlapping open set $U'$ when the transition function is $g(x) = \ln \lambda(x)$.

Now, since only the Lie algebra of $G$ has been defined, namely, the unique one-dimensional Lie algebra (which must then be Abelian) there are two (connected) possibilities for the Lie group $G$: Either it is the multiplicative group $(\mathbb{R}^+, \times)$ of positive real numbers or it is the multiplicative group $U(1)$ of complex numbers of unit modulus, which is also isomorphic to the group of planar Euclidian rotations. Such numbers are usually interpreted as a choice of *phase* in the context of harmonic oscillations, and since the experimental verification the Bohm-Aharonov effect suggests that $U(1)$ was the correct choice, we shall make that choice.

Thus, a transition function will take the form of:

$$g(x) = e^{-i\lambda(x)}, \tag{12.4}$$

and multiplication by this factor becomes the *gauge transformation of the first kind.*

Now, if the 1-form $A$ is a connection 1-form (with values in the Lie algebra of $U(1)$, which is simply the imaginary numbers) on a $U(1)$-principle bundle $\pi: P \rightarrow M$ then the fiber $\pi^{-1}(x)$ over any point $x \in M$ looks like the circle $S^1$, although the diffeomorphism is not unique. In fact, that diffeomorphism usually comes about by making a local choice of $U(1)$ gauge on some open subset $U \subset M$, which then amounts to a local section $\phi: U \rightarrow P$. The part of $P$ that projects onto $U$, which we shall denote by $\pi^{-1}(U)$, then becomes diffeomorphic to $U \times U(1)$ by taking the point $p_x \in \pi^{-1}(x)$ to the point $(x, g(x))$ in such a way that when $p_x = \phi(x)$ the corresponding point in $U \times U(1)$ is $(x, 1)$. More specifically, the element $g(x) \in U(1)$ makes:

$$p_x = g(x)\ \phi(x). \tag{12.5}$$

In effect, the local section $\phi$ becomes equivalent to the function $g: U \rightarrow U(1)$.

---

([1])  This assumes that $M$ is path-connected, which is reasonable for most realistic physical theories. Otherwise, one simply says that there could be a separate constant associated with each path component of $M$.

If the open subset $U \subset M$ carries a coordinate system $x^\mu$, $\mu = 0, 1, \ldots, 3$ and $U(1)$ has the single coordinate $g$ then the local form of $A$ is:

$$A = A_\mu(x, g) \, dx^\mu + g^{-1}dg = A_\mu(x, g) \, dx^\mu + d\lambda \qquad (\lambda = \ln g). \qquad (12.6)$$

One can then pull the 1-form $A$ on $P$ down to a local representative $\phi^*A$ on $U$ by means of the local section $\phi$, and one can then express $\phi^*A$ at each $x \in U$ in the form:

$$A|_x = A_\mu(x) \, dx^\mu, \qquad [A_\mu(x) \equiv A_\mu(x, e)] \qquad (12.7)$$

since the gauge choice $d\lambda = 0$ is associated with $g = 1$.

Now, one of the defining properties of a connection 1-form on any fiber bundle, such as the 1-form $A$ on $P \to M$, is that its annihilating hyperplane $\text{Ann}_p[A]$ at each $p \in P$ must represent a "horizontal" subspace $H_p \subset T_p(P)$, which is complementary to the "vertical" subspace $V_p \subset T_p(P)$, which, in this case, is the line tangent to the circle in the fiber through $p$, and which projects to zero under the differential map $d\pi$. Thus, the algebraic solution to the Pfaff equation:

$$A = 0 \qquad (12.8)$$

is the vector sub-bundle $H(P) \to P$ of $T(P) \to P$ whose fibers are the horizontal subspaces. Such a horizontal subspace can be regarded as a "lifting" of the tangent space to $M$ at $\pi(p)$, and the curvature of the vector bundle $H(P) \to P$ expresses the obstruction to the extension of that lifting to a global lifting.

As for the vectors tangent to the fibers of $P$, which therefore take the local form:

$$X = X^g \, \frac{\partial}{\partial g}, \qquad (12.9)$$

they are annihilated by $A$ only when they are zero to begin with. In fact, one of the defining properties of any connection 1-form on $P$ is that it must define a linear isomorphism of the vertical vector spaces to $P$ with the Lie algebra of the structure group.

We now return the main subject of the present study, namely, the integrability of the Pfaff equation and how it relates to the various mathematical models for physical phenomena. We then see that an integral submanifold of equation (12.8) will be a submanifold $f: N \to P$ such that the tangent spaces to the image $f(N)$ are subspaces of $H_p$ at each point $p$ in that image. Thus, the dimension of $N$ will be at most $4 = \dim(M) = \dim(P) - 1$, although in order to establish that maximum dimension more precisely, one must look at the sequence of integrability forms for $A$:

$$A, \quad dA = F, \quad A \wedge dA = A \wedge F, \quad dA \wedge dA = F \wedge F, \quad A \wedge dA \wedge dA = A \wedge F \wedge F, \ldots$$

of course, when $\dim(M) = 4$, the remaining terms in this sequence vanish.

Since the horizontal subspaces are always complementary to the vertical subspaces, the integral submanifolds of $A$ must be transverse to the fibers of $P$ and the composition of $f$ with the projection $\pi\colon P \to M$ will always define a submanifold of $M$ that has the same dimension as $f$. In other words, an integral submanifold of $A = 0$ is a "horizontal" submanifold of $P$.

If the fibration $P \to M$ is trivial − so $P$ is diffeomorphic to $M \times U(1)$ − then it admits global sections $f\colon M \to P$, which will then be four-dimensional integral submanifolds of $A$, as long as $A$ is the connection 1-form whose horizontal subspaces are all $T_xM$ under that trivialization of $P$. Hence, a global section can exist only if there is a completely integrable connection 1-form on $P$; to show that the existence of such a connection implies the triviality of $P$ is more involved.

In general, what the 2-form $F$ represents in the context of connections on $U(1)$ principle bundles is the *curvature* 2-form for the connection 1-form $A$. Since $A$ takes its values in the Lie algebra of $U(1)$, so does $F$. Thus, one can think of non-vanishing curvature as an obstruction to the integrability of the Pfaff equation that is defined by the connection 1-form $A$.

By assumption, in the context of electromagnetism the very presence of an electromagnetic field in spacetime makes $F$ non-zero. Hence, the only hope for the complete integrability of (12.8) is the vanishing of the Frobenius 3-form $A \wedge F$, but first, we look at the condition for the vanishing of $F \wedge F$.

Basically, the 4-form $F \wedge F$ vanishes iff the 2-form $F$ is decomposable; that is, there must be non-collinear 1-forms $\alpha$ and $\beta$ on $P$ such that:

$$F = \alpha \wedge \beta. \tag{12.10}$$

This also says that such an $F$ has rank two.

One of the most important examples of such electromagnetic fields is given by any electromagnetic field of wave type, although there are other non-wavelike electromagnetic fields of rank two. Indeed, static electric and magnetic fields generally take on that form, at least when one chooses a "time-space splitting" of $T(M)$ into a rank-one temporal sub-bundle and a rank-three complementary spatial sub-bundle so that the concept of "static" is well-defined. In general, the rank-four 2-forms, for which $F \wedge F$ is non-vanishing, take the form of linear superpositions of fields that are due to independent sources. Note that since the condition $F \wedge F = 0$ only defines a quadric hypersurface in each vector space $\Lambda^2_p(P)$, which one calls the *Plücker-Klein quadric*, moreover, the rank-four case is "generic," in the sense that its complement has measure zero.

From what we know about the integrability of $A$, we can also express a rank-two $F$ more specifically in terms of exact 1-forms:

$$F = d\mu \wedge dv. \tag{12.11}$$

The possible 1-forms that are associated with a rank-two $F$ are then of the form:

$$A = \mu \, dv \qquad \text{and} \quad A = d\lambda + \mu \, dv. \tag{12.12}$$

The first case is associated with the vanishing of $A \wedge F$, while the second one is associated with its non-vanishing. Thus, the first case gives completely integrability, while the second case is associated with integral submanifolds of maximum dimension three.

When $F \wedge F$ is non-vanishing, this says that it has rank four, so one can express it in the form:

$$F = d\mu_1 \wedge dv^1 + d\mu_2 \wedge dv^2. \tag{12.13}$$

Thus, when $F \wedge F$ is non-vanishing the Pfaffian equation (12.8) cannot be completely integrable, and the integral submanifolds will have maximum dimension three. The canonical form for $A$ in this case will be:

$$A = \mu_1 \, dv^1 + \mu_2 \, dv^2. \tag{12.14}$$

One can also relate the integrability forms to various topological invariants that are associated with the triviality of non-triviality of the bundle $P \to M$, namely, its Chern classes and Chern-Simon secondary characteristic class.

The 2-form $F$ itself is related to the *first Chern class of P*, which we denote by $c_1[P]$. The Frobenius 3-form $A \wedge F$ takes the form of the *first Chern-Simons secondary characteristic class* of the bundle $P$, while the 4-form $F \wedge F$ is then related to the *second Chern class* $c_2[P]$ of $P$. More precisely, these cocyles are defined by evaluating the differential forms on (differentiable singular) cycles of the same dimension:

$$c_1(P)[z_2] = \frac{1}{2\pi} \int_{z_2} F \, , \qquad c_2(P)[z_4] = \left(\frac{1}{2\pi}\right)^2 \int_{z_4} F \wedge F \, . \tag{12.15}$$

Thus, $c_1[P]$ comes from a 2-form while $c_2[P]$ comes from a 4-form. Note that it is entirely possible for a Chern class to vanish, even when the integrand in (12.15) does not.

Although we have defined both of these characteristic classes by means of the curvature 2-form $F$ of the connection $A$, nevertheless, they are actually independent of that choice and are defined solely by the topology of $P$. This, in turn, is based in the topology of $M$, combined with the topology of the fibers of $P$, which is that of the circle, in this case.

**13. Discussion.** Since the scope of this paper was limited to Pfaffian equations that are defined by the vanishing of a single 1-form, an obvious first direction of expansion of scope would be to the more involved study of Pfaffian systems and their physical applications. Such a system then consists of a finite number of 1-forms $\theta^a$, $a = 1, \ldots, k$ that vanish simultaneously on a given $n$-dimensional manifold $M$. Thus, the contact elements that are defined by the annihilating subspaces of the tangent spaces to $M$ will have codimension $k$; i.e., dimension $n - k$. The conditions for integrability of such systems are more complicated that the ones that we have been dealing with here, and get into the extension of the Cauchy-Kowalewski theorem in partial differential equations to the Cartan-Kähler theorem.

However, the physical applications are still quite fundamental. In particular, one can deal with non-holonomic linear constraints that are defined by more than one linear constraint, the integrability of sections of other jet manifold projections than ones that involve $J^1(M; \mathbb{R})$, anholonomic local frame fields, which one encounters in the study of motion relative to "non-inertial" frames, dislocations and disclinations in plastic media, Einstein's theory of gravitation, and non-Abelian gauge field theories, to name a few examples.

Another intriguing direction to pursue is in the formulation of the calculus of variations when one starts with the virtual work functional, rather than an action functional, a topic that the author has discussed elsewhere in some detail [**19**]. Since that functional is based upon a fundamental 1-form that is exact only when there is an action functional, by allowing it to have inexact contributions, one can also include non-conservative forces and non-holonomic constraints. One might even speculate that the inexact contributions to the exact part might somehow represent "quantum" corrections to a classical variational problem that might provide an alternative to the path-integral formulation of quantum physics.

Hopefully, this present list of physical examples of the more tractable case of a single Pfaffian equation will serve as an inducement to pursue the more advanced topics.